\documentclass{aa}  

\usepackage{graphicx}
\usepackage{txfonts}
\usepackage{multirow}
\usepackage{color}
\usepackage{tablefootnote}
\usepackage{subfigure}

\usepackage{color}

\newcommand{\lya}{\textsc{{\rm Ly}\kern 0.1em$\alpha$}}
\newcommand{\Ly}{\textsc{{\rm Ly}\kern 0.1em$\alpha$}}
\newcommand{\MgI}{\ion{Mg}{I}}
\newcommand{\MgII}{\ion{Mg}{II}}
\newcommand{\FeII}{\ion{Fe}{II}}
\newcommand{\FeIIs}{\ion{Fe}{II}*}
\newcommand{\OII}{[\ion{O}{II}]}
\newcommand{\msun}{M$_{\odot}$}
\newcommand{\mpy}{\msun~yr$^{-1}$} 
\newcommand{\mstar}{$M_{\star} $} 
\newcommand{\flux}{ergs~s$^{-1}$~cm$^{-2}$}

\newcommand{\Nmosaic}{1\,443}		

\newcommand{\NSpec}{315}			
\newcommand{\NOII}{274}				
\newcommand{\NFest}{40}

\newcommand{\zmin}{0.85}  
\newcommand{\zmax}{1.50}
\newcommand{\fluxlimit}{$2\times10^{-18}$ (3 $\sigma$) \flux}

\begin{document} 

	\title{The MUSE Hubble Ultra Deep Field Survey:}
    \subtitle{VII. \ion{Fe}{II}* Emission in Star-Forming Galaxies}
    \titlerunning{MUSE UDF Survey: VII. \ion{Fe}{II}* Emission in Star-Forming Galaxies}
     
 
   \author{Hayley Finley\inst{1}
          \and
          Nicolas Bouch\'e\inst{1}
          \and
          Thierry Contini\inst{1}
          \and
          Mieke Paalvast \inst{2}
          \and
          Leindert Boogaard \inst{2}
          \and
          Michael Maseda \inst{2}
		  \and
          Roland Bacon \inst{3}
		  \and
          Jérémy Blaizot\inst{3}
          \and
          Jarle Brinchmann \inst{1,3}
          \and 
          Benoît Epinat \inst{5}
          \and
		  Anna Feltre\inst{3}          
          \and
          Raffaella Anna Marino \inst{6}
          \and 
          Sowgat Muzahid \inst{2}
          \and 
          Johan Richard \inst{3}
          \and 
          Joop Schaye \inst{1}
          \and
          Anne Verhamme \inst{3,7}
          \and
          Peter M. Weilbacher \inst{8}
          \and 
          Lutz Wisotzki \inst{8}
          }

   \institute{Institut de Recherche en Astrophysique et Plan\'etologie (IRAP), Universit\'e de Toulouse, CNRS, UPS, F-31400 Toulouse, France
   \email{hayley.finley@irap.omp.eu}
   \and
  Leiden Observatory, Leiden University, P.O. Box 9513, 2300 RA Leiden, The Netherlands 
   \and
   CRAL, Observatoire de Lyon, CNRS, Université Lyon 1, 9 Avenue Ch. André, F-69561 Saint Genis Laval Cedex, France 
   \and
   Instituto de Astrof{\'i}sica e Ci{\^e}ncias do Espa{\c{c}}o, Universidade do Porto, CAUP, Rua das Estrelas, PT4150-762 Porto, Portugal 
   \and
   Aix Marseille Univ, CNRS, LAM, Laboratoire d'Astrophysique de Marseille, Marseille, France
   \and
   ETH Zurich, Institute of Astronomy, Wolfgang-Pauli-Str. 27, CH-8093 Z\"urich, Switzerland 
   \and 
   Observatoire de Genève, Université de Genève, 51 Ch. des Maillettes, 1290 Versoix, Switzerland
   \and
   Leibniz-Institut für Astrophysik Potsdam (AIP), An der Sternwarte 16, D-14482 Potsdam, Germany 
    }
  \date{Received / Accepted}

 
  \abstract{
Non-resonant \FeII* ($\lambda2365$, $\lambda2396$, $\lambda2612$, $\lambda2626$) emission can potentially trace galactic winds in emission and provide useful constraints to wind models.
From the $3.15\arcmin \times 3.15\arcmin$ mosaic of the Hubble Ultra Deep Field (UDF) obtained with the  VLT/MUSE integral field spectrograph, we identify a statistical sample of \NFest\ \FeIIs\ emitters and 50 \MgII\ ($\lambda\lambda2796,2803$) emitters from a sample of 271 \OII$\lambda\lambda3726,3729$ emitters with reliable redshifts from $z=\zmin - \zmax$ down to \fluxlimit\ (for \OII), covering the \mstar\ range from $10^8 - 10^{11}$ \msun. 
The \FeIIs\ and \MgII\ emitters follow the galaxy main sequence, but with a clear dichotomy.
Galaxies with masses below $10^9$ \msun\ and star formation rates (SFRs) of $\lesssim1$ \mpy\ have \MgII\ emission without accompanying \FeIIs\ emission, whereas galaxies with masses above $10^{10}$ \msun\ and SFRs $\gtrsim 10$ \mpy\ have \FeIIs\ emission without accompanying \MgII\ emission.  Between these two regimes, galaxies have both \MgII\ and \FeIIs\ emission, typically with \MgII\ P-Cygni profiles.
Indeed, the \MgII\ profile shows a progression along the main sequence from pure emission to P-Cygni profiles to strong absorption, due to resonant trapping.
Combining the deep MUSE data with HST ancillary information, we find that galaxies with pure \MgII\ emission profiles have lower star formation rate surface densities than those with either \MgII\ P-Cygni profiles or \FeIIs\ emission.
These spectral signatures produced through continuum scattering and fluorescence, \MgII\ P-Cygni profiles and \FeIIs\ emission, are better candidates for tracing galactic outflows than pure \MgII\ emission, which may originate from \ion{H}{ii} regions.
We compare the absorption and emission rest-frame equivalent widths for pairs of \FeII\ transitions to predictions from outflow models and find that the observations consistently have less total re-emission than absorption, suggesting either dust extinction or non-isotropic outflow geometries.
}

   \maketitle	
%

\section{Introduction}

Galactic winds, driven by the collective effect of hot stars and supernovae explosions,  appear ubiquitous 
 \citep[e.g.,][]{VeilleuxS_05a, WeinerB_09a, SteidelC_10a,RubinK_10a, RubinK_14a,ErbD_12a, MartinC_12a,NewmanS_12a,HarikaneY_14a,BordoloiR_14a, HeckmanT_15a, ZhuG_15a, ChisholmJ_15a},
and are thought to play a major role in regulating the amount of baryons in galaxies  \citep{2012RAA....12..917S}, in enriching the intergalactic medium with metals  \citep{2008MNRAS.387..577O, 2016MNRAS.459.1745F} and in regulating the mass-metallicity relation \citep{2001ApJ...561..521A, 2008MNRAS.385.2181F, 2013ApJ...772..119L,TremontiC_04a}. 
Most studies of galactic winds beyond the local universe rely on detecting low-ionization transitions, like \ion{Si}{II}, \MgII, or NaD, in absorption against the galaxy continuum that have an asymmetric, blue-shifted line profile indicative of outflowing gas.

Another technique for studying galactic winds relies on detecting emission signatures. 
Traditionally, emission signatures used to characterized galactic winds in local ultraluminous infra-red galaxies, are  broad components in optical lines 
 \citep[e.g.,][]{LehnertM_1995a,LehnertM_1996a,VeilleuxS_03a,StricklandD_04a,WestmoquetteM_12a,SotoK_12a,RupkeD_13a,ArribasS_14a}, 
or line ratios diagnostics that indicate shocks, \citep[e.g.][]{VeilleuxS_03a,SotoK_12a}.
Broad H$\alpha$ components from galactic winds can also be detected  in distant $z \approx 2$ star-forming galaxies \citep[e.g.][]{GenzelR_11a,NewmanS_12a}.
Galactic winds are also traced with X-ray emission from shocked gas in local starbursts \citep[e.g.][]{MartinC_99a,LehnertM_99a,StricklandD_99a,StricklandD_04a,StricklandD_09a,GrimesJ_2005a}.
Observing galactic winds directly in emission is nonetheless inherently difficult, because emission processes tend to depend on the square of the gas density and hence have very low surface brightnesses.

\begin{figure*}
\centering
\includegraphics[]{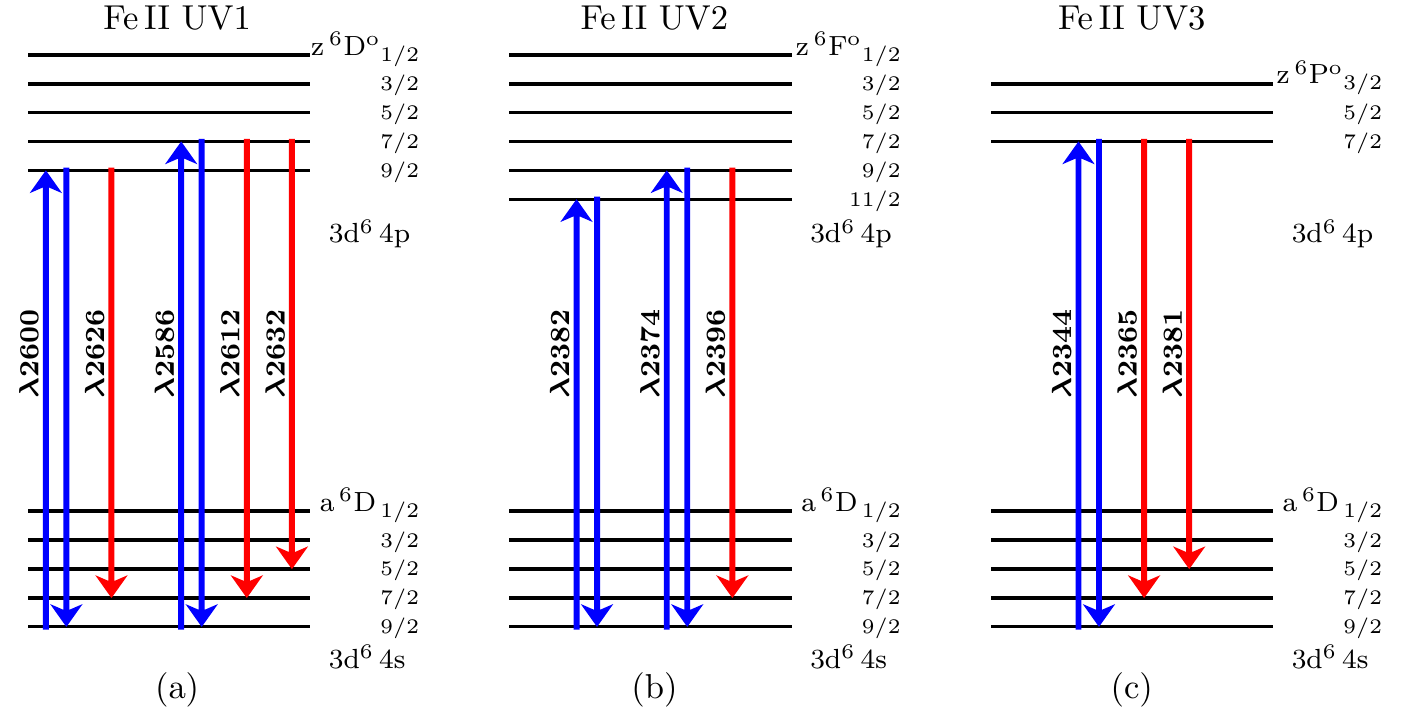}
\caption{Energy level diagrams for the \FeII\ multiplets, UV1 (a), UV2 (b), and UV3 (c), where the ground and the excited states have multiple levels due to fine-structure splitting.
Resonant transitions are shown in blue, and non-resonant transitions are shown in red.
Whether non-resonant emission is likely to occur depends on the de-excitation rates and on the number (0, 1, or 2) of potential re-emission channels \citep{TangY_14a,ZhuG_15a}.
For example, the \FeII\,$\lambda2382$ transition from the UV2 multiplet has no associated \FeIIs\ emission lines, and thus behaves like a purely resonant transition (e.g., \lya\ or \MgII).
}\label{fig:EnergyDiagrams}
\end{figure*}

A relatively new technique for studying galactic winds in emission relies on studying the signatures of photon scattering in low-ionization transitions 
since the pioneering work of \citet{RubinK_11a}. 
Photons absorbed in low-ionization metal lines  (e.g., \ion{Si}{II}, \ion{C}{II}, \FeII, \MgII)   can then lead to resonant or non-resonant re-emission.  
For resonant transitions, re-emitting absorbed photons through the same transition can give rise to P-Cygni profiles with blue-shifted absorption and redshifted emission depending on the line optical depth, geometric factors, and the amount of emission infilling, as discussed in \citet{ProchaskaJ_11a}. 
For non-resonant transitions, which are commonly indicated with an asterisk (e.g., \ion{Si}{II}*, \ion{C}{II}*, and \FeIIs), resonantly absorbed photons are re-emitted to one of the split levels of the ground state (e.g., Fig.~\ref{fig:EnergyDiagrams}).
The resulting non-resonant emission lines, produced through continuum fluorescence, are typically a few Angstroms redward of their originating absorption lines. 
Resonant \MgII\ ($\lambda\lambda2796,2803$) emission and non-resonant \FeII* ($\lambda2365$, $\lambda2396$, $\lambda2612$, $\lambda2626$) emission were first recognized as potential signatures of galactic winds in emission when seen together in the spectrum of a $z = 0.694$ star-forming galaxy \citep{RubinK_11a}.

Characterizing the properties of galaxies that exhibit \FeIIs\ and \MgII\ emission, typically with corresponding \FeII\ and \MgII\ absorption, is important for understanding the physical conditions that lead to outflows. 
Since \FeII\ and \MgII\ have similar ionization potentials, 7.90~eV and 7.65~eV respectively \citep[NIST-ASD database; see also Table 2 from][]{ZhuG_15a}, they trace the same gas phase in the outflows.
Galaxy properties, such as dust content, gas density, and inclination (for non-isotropic outflows), modulate the amount of resonant and non-resonant emission predicted in radiative transfer models of galactic outflows \citep{ProchaskaJ_11a,ScarlataC_15a}.
In the local universe, studies focused on resonant \ion{Na}{i}~D absorption and emission, which behave like \MgII, have been able to investigate the connection between galaxy properties and outflows by leveraging a large statistical sample to trace, for example, how the emission and absorption varies with galaxy inclination \citep{2010AJ....140..445C} or spatially resolving the emitting region for an individual galaxy \citep{2015ApJ...801..126R}.

Similar analyses for galaxies that exhibit \FeIIs\ and \MgII\ emission are limited, because individual detections of non-resonant \FeII* emission exist for only a handful of $z \lesssim 1$ galaxies \citep[e.g.][]{RubinK_11a, CoilA_11a, MartinC_12a,FinleyH_17a}.
For instance, \citet{FinleyH_17a} found that the \FeIIs\ spatial extent is 70\%\ larger than that of the stellar continuum emission for an individual $z=1.29$ galaxy observed with the Multi-Unit Spectroscopic Explorer \citep[MUSE;][]{BaconR_15a} instrument.
Such individual detections of non-resonant \FeII* emission are rare, because slit losses may preclude detecting \FeII* emission with traditional spectroscopy \citep{ErbD_12a, KorneiK_13a, ScarlataC_15a}.
The MUSE integral field unit instrument eliminates the problem of slit losses and also offers a substantial gain in sensitivity, with a throughput of 35\% end-to-end including atmosphere and telescope at 7000~\AA.

Since direct detections of individual galaxies with signatures of outflows in emission are difficult, several studies have instead focused on characterizing \FeII* and \MgII\ emission by creating composite spectra from $\sim 100$ or more $z \sim 1$ star-forming galaxies \citep{ErbD_12a, KorneiK_13a, TangY_14a,ZhuG_15a}.
These studies then look for trends between the emission strength and galaxy properties, such as stellar mass or dust extinction, by making composite spectra from sub-samples of galaxies.
\citet{ErbD_12a} find that the most striking difference is between low and high-mass galaxies (median stellar masses of $1.8\times 10^{9}$~\msun\ and $1.5\times 10^{10}$~\msun, respectively) with both stronger \MgII\ emission and stronger \FeIIs\ emission in the low-mass composite spectrum.  
%
Interestingly, \citep{ErbD_12a} find more \FeII* emission for galaxies with strong \MgII\ emission. 

After testing the emission strengths in 18 sets of composite spectra, \citet{KorneiK_13a} argue that dust extinction is the most important property influencing \FeII* emission and is also a key property promoting \MgII\ emission (more emission for lower dust extinction in both cases). 
\citet{KorneiK_13a} also find that galaxies with higher specific star-formation rates (sSFR) and lower stellar masses have stronger \MgII\ emission, whereas galaxies with lower star formation rates (SFR) and larger \OII\ equivalent width measurements ($W_{\OII}$) have stronger \FeII* emission. 

Unlike the two previous studies, \citet{TangY_14a} do not find any strong trends with stellar mass, SFR, sSFR, or E($B-V$). \citet{TangY_14a} focus only on the \FeII* emission and associated \FeII\ absorption properties.
Nonetheless, in an analysis of 8\,620 emission-line galaxies, \citet{ZhuG_15a} find that \FeII* emission strength increases almost linearly with $W_{\OII}$.   

A major caveat is that stacking offers little insight into how the emission might depend on wind orientation or geometry given that composite spectra average out all galaxy inclinations.
These geometrical effects can potentially be important, as radiative transfer models of outflows demonstrate \citep[i.e.,][]{ProchaskaJ_11a,ScarlataC_15a}.
Characterizing how geometrical effects impact the emission signatures of outflows can only be performed with a sample of individual galaxies.

Thanks to the recent deep observations of the {\it Hubble} Ultra Deep Field South (UDF) with MUSE 
\citep[][, hereafter Paper I]{Bacon2017}, we can now study and characterize a statistical sample of individual (unlensed) galaxies with \FeIIs\ in emission in order to understand whether geometrical effects play a role in \FeIIs\ emission (and/or \MgII\ emission). 
We can also investigate how the prevalence of \FeIIs\ non-resonant emission varies with galaxy properties such as stellar mass, (specific) star formation rate, etc., thanks to deep multi-band photometry in the $3.15\arcmin \times 3.15\arcmin$ mosaic of the UDF. This paper focuses on the emission line properties, and we will present the absorption line analysis and kinematics in a forthcoming paper.

The paper is organized as follows. In section \S~\ref{sec:data}, we present the data and our selection criteria for \FeIIs\ emitters (and \MgII\ emitters).
In section \S~\ref{sec:results}, we present our main results regarding the statistical properties of \FeIIs\ emitters.
In section \S~\ref{sect:representative}, we show five representative cases.
Finally, we review our findings in Sect.~\ref{sec:conclusions}.
Throughout the paper, we assume a $\Lambda$CDM cosmology with $\Omega_{\rm m} = 0.3$, $\Omega_{\Lambda} = 0.7$, and $H_0 = 70$~km~s$^{-1}$~Mpc$^{-1}$.

\section{Data}
\label{sec:data}

\subsection{MUSE Observations}
We used the $3.15\arcmin \times 3.15\arcmin$ mosaic observations from nine MUSE pointings of the Hubble Ultra Deep Field South presented in Paper I. In summary, the MUSE UDF was observed during eight GTO runs over two years, from September 2014 to December 2015, for a total of 227 25-minute exposures, leading to a depth of $\sim$10~hours per pointing. The central pointing (referred to as UDF-10) was observed for an additional 20~hours, leading to a total depth of $\sim$30~hours in this region.
The median PSF is 0.6\arcsec, and the final 10-hour data cube reaches a depth
of $\sim$ \fluxlimit\ for line emitters (point sources).
Further details about the observations and data reduction are presented in Paper~I.

We used the MUSE UDF redshift catalog presented in \citet{Inami2017} (paper~II). The paper~II authors first identified sources in the MUSE data cube from objects with F775W$\leq27$~mag in the UVUDF photometric catalog \citep{RafelskiM_15a} and from a blind search for emission lines objects using the ORIGIN software (Mary et al., in prep).
The paper~II authors then combined a modified version of the AUTOZ \citep{Baldry_14a} cross-correlation algorithm with the MARZ software \citet{Hinton_16a} to determine the redshifts.
While verifying the algorithm results, the paper~II authors assigned a confidence level (CONFID) from 1 to 3 to each redshift measurement, where CONFID~$=$~1 corresponds to the lowest confidence measurements and CONFID~$=$~3 indicates the highest confidence measurements based on the presence of multiple absorption or emission features.
They measured redshifts for \Nmosaic\ objects in the $3.15\arcmin \times 3.15\arcmin$ MUSE UDF mosaic, of which 196, 684 and 563 objects have redshift confidence 1, 2 and 3, respectively. Secure redshift measurements have CONFID~$>$~1. 

\subsection{Sample selection}
\label{sec:sample}

Since \citet{FinleyH_17a} demonstrated the advantages of detecting \FeIIs\ from an individual galaxy, we took the MUSE UDF mosaic catalog (Paper II) as a basis to build a statistically significant sample of galaxies with \FeII* emission/outflow signatures. As described in the previous section, this catalog includes \Nmosaic\ objects with measured redshifts from an area of $3.15\arcmin \times 3.15\arcmin$ observed to a depth of \fluxlimit\ in $\sim10$~hours.
Using this catalog, we first imposed a redshift range $\zmin-\zmax$ designed such that we cover at least the \OII\,$\lambda\lambda3727,3729$ line and the UV1~\FeII\ multiplet, including the \FeIIs\ emission lines at $\lambda2612$ and $\lambda2626$. 
Although the MUSE spectral coverage for \FeIIs\ extends beyond $z = \zmax$, 
this upper limit ensures covering the \OII\ nebular line, which provides reliable systemic redshifts and a standardized approach to determining star-formation rates.

From the UDF mosaic catalog of \Nmosaic\ objects with measured redshifts, \NSpec\ galaxies are in the redshift range $\zmin-\zmax$. From these \NSpec\ galaxies, we kept \NOII\ galaxies with redshift confidence CONFID~$>$~1, of which 234 (40) have redshift confidence 3 (2), respectively. All but three of these galaxies are \OII\ emitters.

Within this sample, we visually inspected the spectra and searched for signatures of \FeIIs.
We flagged a galaxy as an \FeIIs\ emitter if the spectrum shows any \FeIIs\ emission at $\lambda2612$ and $\lambda2626$ from the UV1 multiplet, at $\lambda2396$ from the UV2 multiplet, or at $\lambda2365$ from the UV3 multiplet, if covered.\footnote{In the MUSE UDF spectra, we do not detect \FeIIs emission at $\lambda2381$ or $\lambda2632$. The \FeIIs\,$\lambda2381$ transition is blended with the \FeII\,$\lambda2382$ absorption.  }
Similar to the CONFID flag in the UDF mosaic catalog, we applied a quality control (qc) flag during the visual inspection. 
The qc > 1 flag indicates spectra with at least two \FeIIs\ emission lines (secure detections), whereas q = 1 designates more marginal cases.
As summarized in Table~\ref{table:sample_numbers}, we found \NFest\ \FeIIs\ emitters in the UDF mosaic, 25 of which have qc~$> 1$. 
All of the galaxies with \FeIIs\ emission also have \FeII\ absorption.

In order to investigate the \MgII\ emission properties of galaxies from the same parent sample and compare them with the \FeIIs\ emission properties, we simultaneously flagged the \MgII\ profiles of the \NOII\ galaxies in our redshift range as pure emission, P-Cygni or pure absorption. 
The \MgII\,$\lambda\lambda2796,2803$ doublet is always covered within the $\zmin-\zmax$ redshift range. 
In the UDF mosaic, we found 33 galaxies with pure \MgII\ emission and 17 galaxies with P-Cygni profiles.

\begin{table} 
\caption{UDF mosaic outflow signature galaxy sample}  
\label{table:sample_numbers}
\centering                                      
\begin{tabular}{l c  c}          
\hline\hline   
    Spectral Signature & Total~~ & qc~$> 1$  \\
\hline             
\OII\ emitters & 271~~  & -- \\
\FeIIs\ emitters & 40 & 25 \\
\MgII\ emitters & 33 & 20 \\
\MgII\ P-Cygni & 17 & 13 \\
\MgII\ absorbers & 40 & 29 \\
\FeII\ absorbers & 72 & 59 \\
\hline
\end{tabular}
\end{table}

\section{Results for \FeIIs\ and \MgII\ emitters}
\label{sec:results}

\subsection{Redshift dependence of \FeIIs\ and \MgII\ emitter fractions}
\label{sec:zdepend}

\begin{figure*}[!t]
\centering
\subfigure[]{\includegraphics[width=8cm]{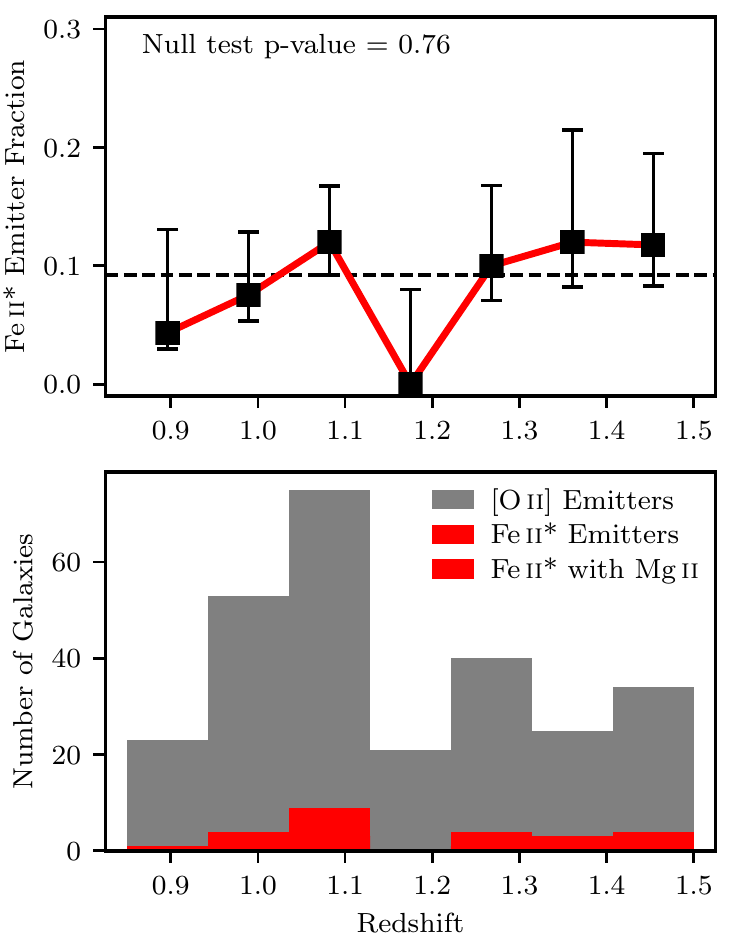}}
\hspace{0.5cm}
\subfigure[]{\includegraphics[width=8cm]{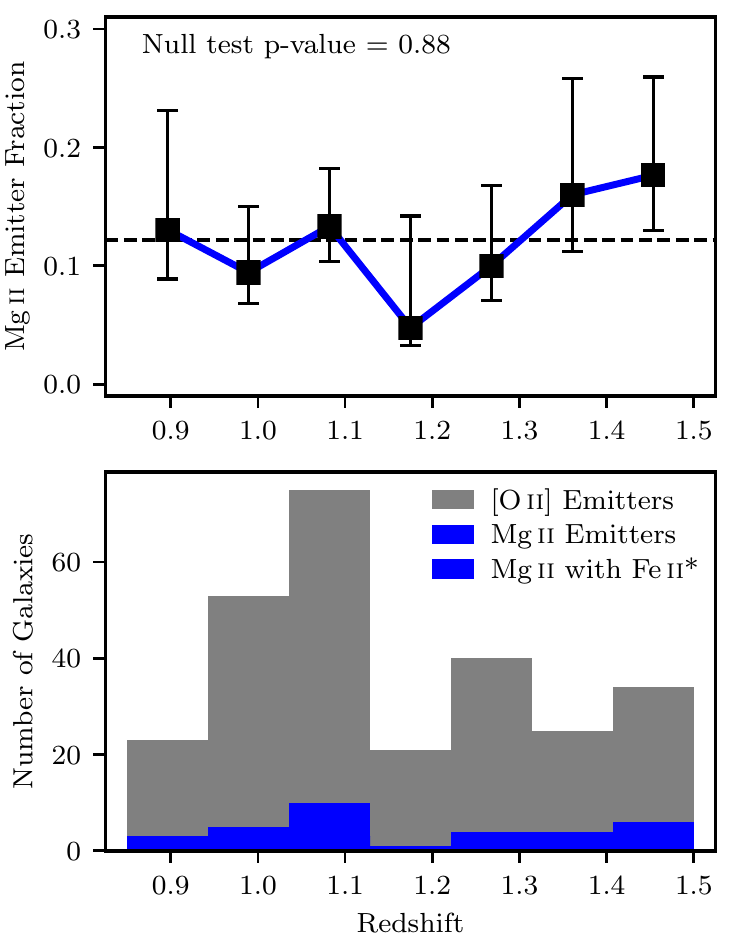}}
\caption{\textbf{a)}: {\it Bottom:} Redshift distribution for the \FeIIs\ emitters. 
The grey histogram shows the distribution for the full sample of 271 \OII\ emitters in the redshift range $\zmin < z < \zmax$ (271 galaxies), and the red histogram shows the subpopulation of \FeIIs\ emitters with confidence flag qc~$> 1$ (25 galaxies). White hatching indicates \FeIIs\ emitters that also have \MgII\ emission or P-Cygni profiles (9 galaxies).
{\it Top:} The fraction of \FeIIs\ emitters for the eight redshift bins. Error bars on these fractions represent 68\%\ confidence levels using Beta distributions as in \citet{CameronE_11a}. The \FeIIs-emitter fraction is about $10$\%\ globally and is also consistent with a uniform distribution.  
\textbf{b)}: {\it Bottom:} Redshift distribution for the \MgII\ emitters. 
The grey histogram again shows the distribution for the full sample of \OII\ emitter galaxies, and the blue histogram shows the subpopulation of \MgII\ emitters with confidence flag qc~$> 1$ (33 galaxies). White hatching indicates \MgII\ emitters that also have \FeIIs\ emission (9 galaxies).
{\it Top:} The fraction of \MgII\ emitters for each redshift bin with 68\%\ confidence intervals. The \MgII-emitter fraction is about $12$\%\ globally and is also consistent with a uniform distribution.  
}
\label{fig:zDist}
\end{figure*}
 
We first look at the redshift distribution of our \FeIIs\ emitters to check whether they occur at a preferred redshift compared to the parent population of emission-line selected \OII\ emitters. The \OII\ emitters have a flux distribution that is approximately constant with redshift.
\footnote{The parent population of \OII\ emitter galaxies appears non-uniform, since skyline emission at redder wavelengths interferes with our ability to detect \OII\ emitters towards higher redshifts. See \citet{Brinchmann2017} for a discussion of redshift completeness in the MUSE UDF catalog.}.

We can expect that the redshift distribution will show a uniform relative fraction of \FeIIs\ emitters, if galactic outflows are ubiquitous in star-forming galaxies.
However, \cite{KorneiK_13a} found that higher redshift galaxies have stronger \FeIIs\ emission in composite spectra from a sample of 212 star-forming galaxies with $0.2 < z < 1.3$ ($\langle z \rangle = 0.99$), which the authors suggest could be due to galaxy properties evolving with redshift. If higher redshift galaxies produce stronger \FeIIs\ emission, then potentially we would detect more \FeIIs\ emitters at higher redshift. 

Fig. ~\ref{fig:zDist}(a) traces the redshift distribution of galaxies across the range $\zmin < z < \zmax$.
In the bottom panel, the grey histogram shows the parent sample of 271 \OII\ emitter galaxies, and the red histogram shows the \FeIIs\ emitters.
The top panel plots the fraction of \FeIIs\ emitters in each redshift bin with error bars representing the 68\%\ confidence interval calculated from the Beta distribution following \citet{CameronE_11a}.
On average across the redshift range, the fraction of \FeIIs\ emitters is $\sim10$\%.

We test the observed fraction of \FeIIs\ emitters against the null hypothesis of  a constant fraction over the redshift range using the proportions $\chi^2$ test from the Python statmodels module.\footnote{Through MonteCarlo testing, we verified that the proportions $\chi^2$ follows a $\chi^2$ distribution even in the low-count regime, unlike the Pearson $\chi^2$.}  
Based on the p-value of 0.40, the fraction of \FeIIs\ emitters does not show evidence of evolving across the redshift range $\zmin < z < \zmax$.
Since our redshift range does not extend to as low redshifts as the \citet{KorneiK_13a} sample, we may not be as sensitive to the effects of galaxy evolution that could produce less \FeIIs\ emission at lower redshift.

Similarly, Fig.~\ref{fig:zDist}(b) compares the redshift distribution of galaxies with pure \MgII\ emission to the parent sample of \OII\ emitters. 
Based on applying the $\chi^2$ test, the relative fraction of \MgII\ emitters also does not evolve with redshift across the redshift range $\zmin < z < \zmax$. The average fraction is $\sim11$\%, comparable to the average fraction of \FeIIs\ emitters. 

The redshift distributions for the \FeIIs\ and the \MgII\ emitters are similar. We applied a Kolmogorov-Smirnov (KS) test to compare the redshift distributions for the galaxies with only \FeIIs\ emission and only \MgII\ emission (excluding galaxies with both \FeIIs\ and \MgII\ emission). The KS test results in a p-value of 0.79, suggesting that these two independent populations could be drawn from the same distribution.
The phenomena producing \FeIIs\ and \MgII\ emission occur in 18\% of star-forming galaxies ($49/271$) observed in the MUSE UDF with a uniform distribution across the redshift range $\zmin < z < \zmax$. 


\begin{figure*}[!ht]
\centering
\subfigure[from SED fitting with FAST]{
\includegraphics[width=8.8cm]{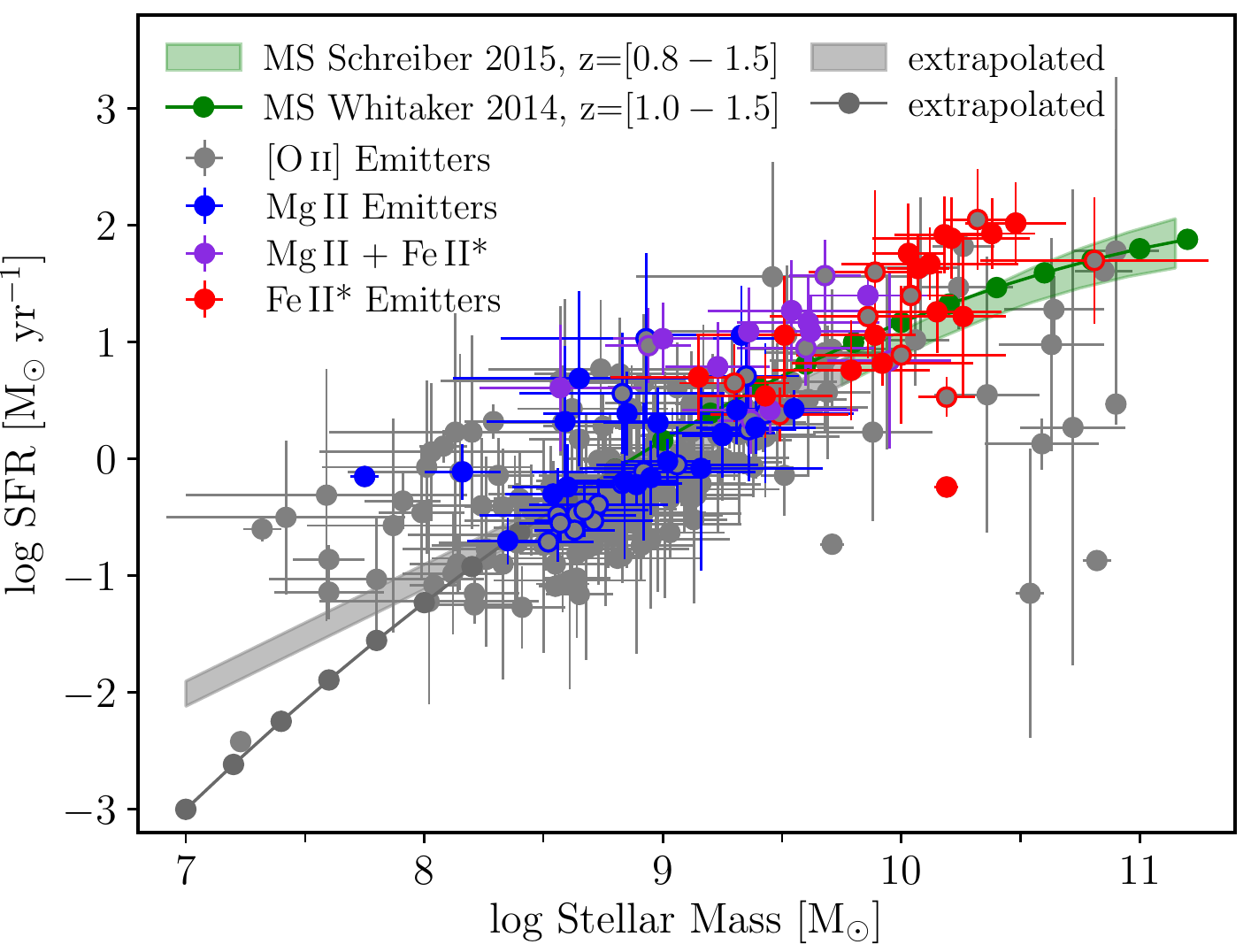}
}
\subfigure[from L\OII\ with a dust correction]{
\includegraphics[width=8.8cm]{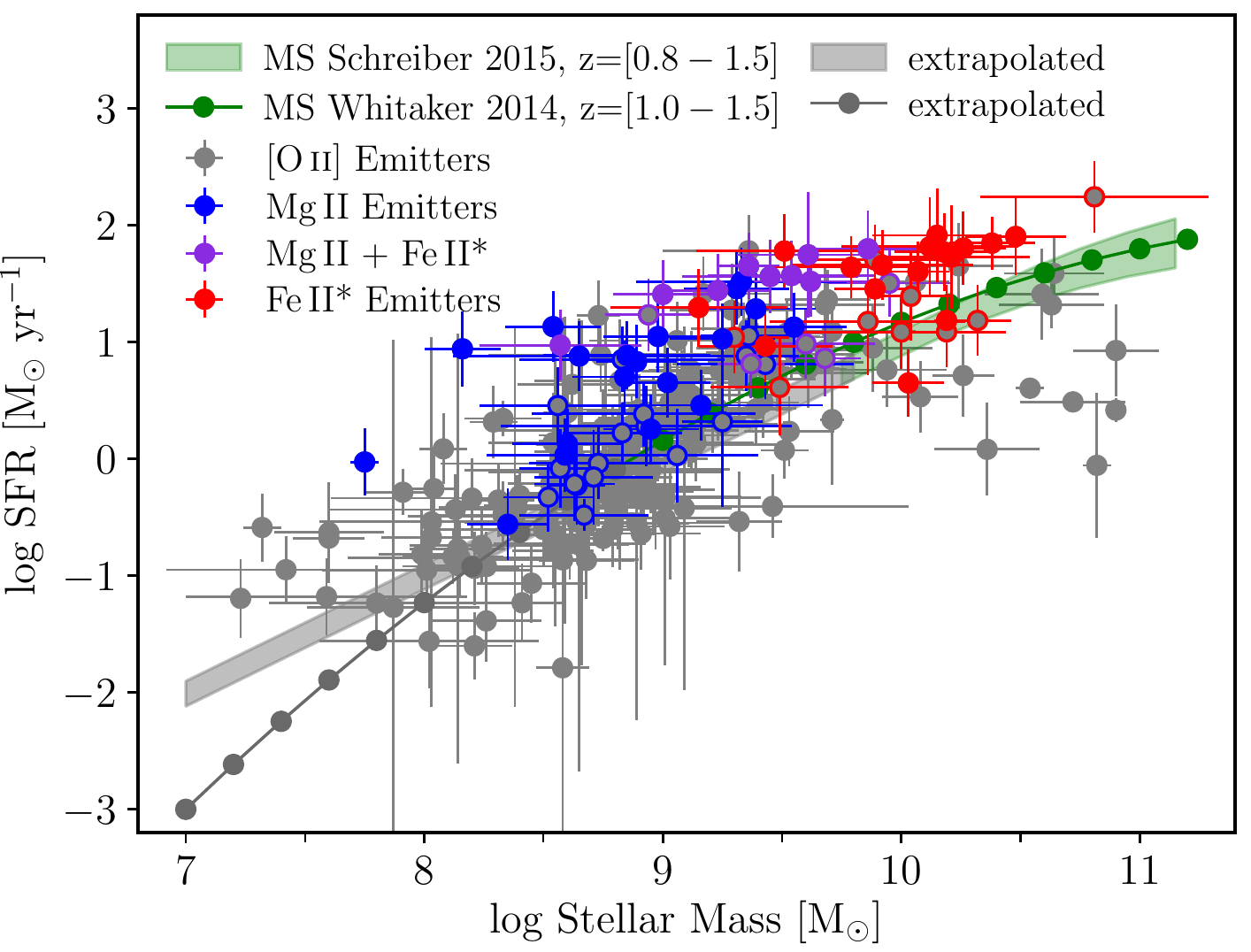}
}
\caption{{\bf a)}: SFR--\mstar sequence for the 271 galaxies in our redshift range, ($\zmin < z < \zmax$), using SFR values from SED fitting.
{\bf b):} SFR--\mstar sequence for the same galaxy sample using SFR values from L\OII\ fluxes with a dust correction following \citet{KewleyL_04a}. 
In both panels, galaxies with only \FeIIs\ emission (only \MgII\ emission or P-Cygni profiles) are shown in red (blue).
Galaxies with both \FeIIs\ emission and \MgII\ emission or P-Cygni profiles are shown in purple.
Filled colored points indicate secure detections with qc~$> 1$, and points with colored outlines indicate qc~$= 1$ detections.
The green filled region represents the main sequence in our redshift range determined by \citet{SchreiberC_15a} using a mass complete sample of 60\,000 galaxies from the GOODS-{\it Herschel} and CANDELS-{\it Herschel} programs. The grey filled region represents the main sequence from \citet{SchreiberC_15a} extrapolated below their mass completeness.
The green (grey) solid line with circular points represents the main sequence from \citet{WhitakerK_14a} over the redshift range $z=[1.0-1.5]$ to \mstar$=10^9$ \msun\ (extrapolated below their completeness), respectively. 
}
\label{fig:Fest:MS}
\end{figure*}

\subsection{\FeIIs\ and \MgII\ emitters on the Main Sequence}  
\label{sec:mainseq}

We now turn towards the galaxy star-formation main sequence. 
This scaling relation between star-formation rate (SFR) and \mstar\ is particularly important \citep{BoucheN_10a,MitraS_17a}, 
since it applies for star-forming galaxies from the local universe to $z \gtrsim 4$. 
Based on the work of numerous authors \citep[e.g.,][among the more recent surveys]{KarimA_11a,WhitakerK_14a,SchreiberC_15a}, the galaxy main sequence is almost linear except perhaps for \mstar~$>10^{10}$~\msun.
Depending on where the \FeIIs\ and \MgII\ emitters fall on this relation, the galaxy main sequence allows us to identify whether they are typical star-forming galaxies or if they instead belong to a subpopulation, such as starburst galaxies.


In order to estimate the stellar masses of the galaxies in the MUSE mosaic catalog, we performed standard
spectral energy distribution (SED) fitting to the {\it HST} ACS and WFC3 photometry.
We followed the same procedure as in Boogaard et al.\ (in prep) and Paalvast et al.\ (in prep). 
Briefly, this procedure applies the FAST (Fitting and Assessment of Synthetic Templates) algorithm \citep{KriekM_09a} using the 10 {\it HST} filters from \citet{RafelskiM_15a} and the \citet{BruzualG_03a} library.  We assumed exponential declining star formation histories  with a \citet{CalzettiD_00a} attenuation law and a \citet{ChabrierG_03a} initial mass function (IMF).

As described in the section~\ref{sec:sample}, we selected galaxies with a maximum redshift \zmax, thereby ensuring that we cover \OII. We estimated the \OII-based SFRs from the luminosity $L_{\OII,\rm  obs}$ using the method described in \citet{KewleyL_04a}, which includes an empirical dust correction (their Eq. 17 \& 18) and a metallicity correction (their Eq.10 or 15). The metallicity $Z$ is estimated from the \mstar--Z relation of \citet{ZahidJ_14b} and their formalism. 
To make the underlying \citet{SalpeterE_55a} IMF for the \OII-based SFRs consistent with the \citet{ChabrierG_03a} IMF used for the SED-based SFRs, we divided the \OII-based SFRs by a factor of $1.7$.


The left (right) panel in Fig.~\ref{fig:Fest:MS} shows the SFR main sequence for our sample using SFR values from SED modeling ($L_{\rm \OII}$ nebular models), which produce overall consistent main sequences.  
Fig.~\ref{fig:Fest:MS} also indicates the main sequence that \citet{SchreiberC_15a} determined from a sample of 60\,000 galaxies (mass complete down to $\sim 10^{9.8}$ \msun) from the GOODS-{\it Herschel} and CANDELS-{\it Herschel} key-programs (green filled region) and that \citet{WhitakerK_14a} found for the redshift range $z=[1.0-1.5]$ to \mstar$=10^9$ \msun\ (green solid line with filled points). 
We extrapolated the results from \citet{SchreiberC_15a} and \citet{WhitakerK_14a} below their mass completeness to better compare with our sample (gray filled region and dark gray solid line with filled points, respectively).
The UDF mosaic galaxies follow the expected trends down to $\sim 10^{8}$ \msun. (See also Boogaard et al.\ (in preparation) for a discussion of the main sequence properties at the low-mass end.)

In Fig.~\ref{fig:Fest:MS}, grey points indicate galaxies from our sample that have \OII\ emission, but no \FeIIs\ or \MgII\ emission.
Red (blue) points represent galaxies with only \FeIIs\ emission (only \MgII\ emission), whereas purple points indicate galaxies that have both \FeIIs\ emission and \MgII\ emission. Here we include galaxies with P-Cygni profiles in the \MgII\ emitter sample. 
This figure reveals that there is a strong apparent dichotomy between the populations of \FeIIs\ and \MgII\ emitters.
Indeed, below $10^9$ \msun\ (and SFRs of $\lesssim1$ \mpy), we observe \MgII\ emission without accompanying \FeIIs\ emission, whereas, above $10^{10}$ \msun\ (and SFRs $\gtrsim 10$ \mpy), we observe \FeIIs\ emission without accompanying \MgII\ emission.  Between these two regimes, we observe both \MgII\ and \FeIIs\ emission, typically with \MgII\ P-Cygni profiles.


The dichotomy between \MgII\ and \FeIIs\ emitters shown in Fig.~\ref{fig:Fest:MS} could be the result of a selection effect due to different sensitivities for \MgII\ and \FeIIs\ in the spectra. Two potential selection effects could affect our sample, one that would prevent us from observing \MgII\ emission in high-mass galaxies and another that would prevent us from detecting \FeIIs\ emission in low-mass galaxies. The first selection effect can be ruled out, because the spectra with the largest signal-to-noise are for galaxies with strong continua, typically at high-masses. Moreover, the ability to detect a constant flux/equivalent width does not depend on the continuum strength. 

The second selection effect could explain
the lack of \FeIIs\ emission at low mass and low SFR, because we need greater sensitivity in order to detect the \FeIIs\ emission, which is inherently weaker. Indeed, the strongest \FeIIs\ emission lines typically have rest-frame equivalent widths W$_0$ between $-0.5$ and $ -1~\AA$, whereas the \MgII\ emission lines have rest-frame equivalent widths $-1$ and $-5~\AA$ (See Feltre et al.\ (in preparation) for \MgII\ emission properties.)
Examining the 30-hour spectra from \MgII\ emitters in the UDF-10, only one reveals \FeIIs\ emission and \FeII\ absorption that were not flagged in the 10-hour spectra (Sect.~\ref{sect:representative}).
However, even if we miss accompanying \FeIIs\ emission for the low-mass \MgII\ emitters, we still observe a progression in \MgII\ spectral signatures along the main sequence. We discuss physically motivated reasons for the \MgII\ and \FeIIs\ spectral signatures in Sect.~\ref{discussion}.
%
%



An important caveat to comparing the \MgII/\FeIIs\ dichotomy in Fig.~\ref{fig:Fest:MS} with trends from composite spectra is that the samples used to create the composite spectra have almost no galaxies with \mstar~$= 10^{8-9}$~\msun\ and SFR~$< 1$~\mpy, the regime where we observe \MgII\ emission without accompanying \FeIIs\ emission. 
The composite spectra are only sensitive to the \mstar--SFR regime where we observe \FeIIs\ emission from the individual MUSE galaxies. Indeed, the regime that their sample covers may explain why \citet{TangY_14a} do not see strong differences in the \FeIIs\ emission from their composite spectra split by stellar mass or SFR. 
Both \citet{ErbD_12a} and \citet{KorneiK_13a} find that composite spectra with strong \MgII\ emission also have strong \FeIIs\ emission.
Similar to many of the individual MUSE UDF galaxies with \mstar~$\sim 10^{9.5}$~\msun, such as Fig.~\ref{fig:udf10-id32}, these composite spectra show \FeIIs\ emission and \MgII\ P-Cygni profiles.
Again, the \mstar--SFR regimes that the composite spectra studies probe implies that they are comparing samples of galaxies where we observe both \MgII\ and \FeIIs\ emission from the MUSE galaxies.

\subsection{\FeIIs\ and \MgII\ emission as a function of galaxy inclination and size}

\begin{figure*}[!t]
\centering
\subfigure[][]{
\includegraphics[width=8cm]{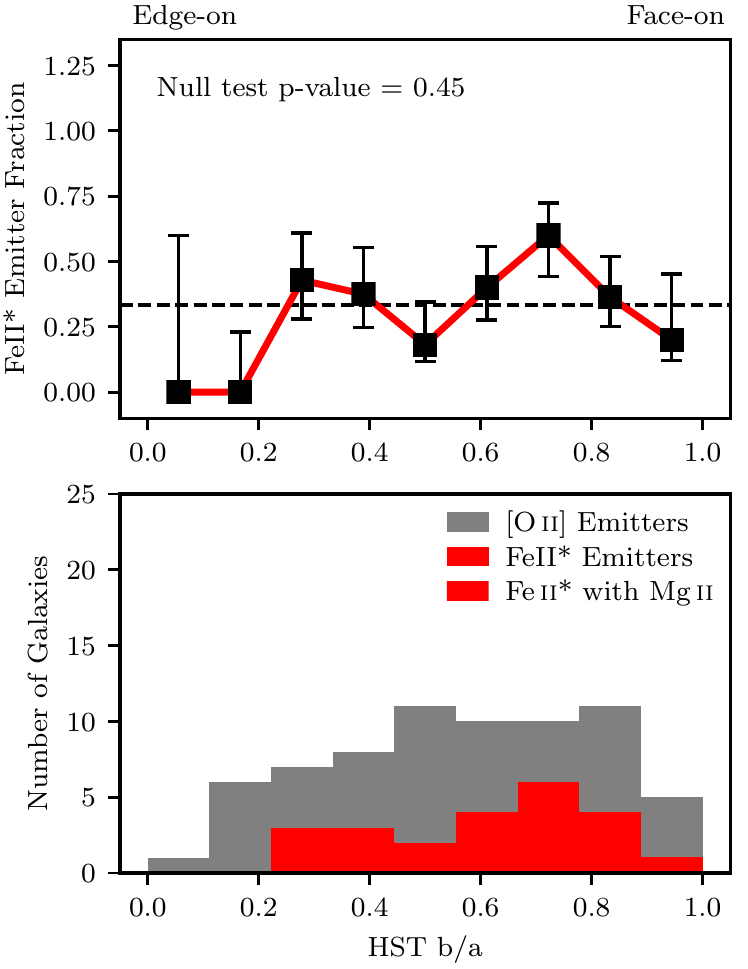}}
\hspace{0.3cm}
\subfigure[][]{
\includegraphics[width=8cm]{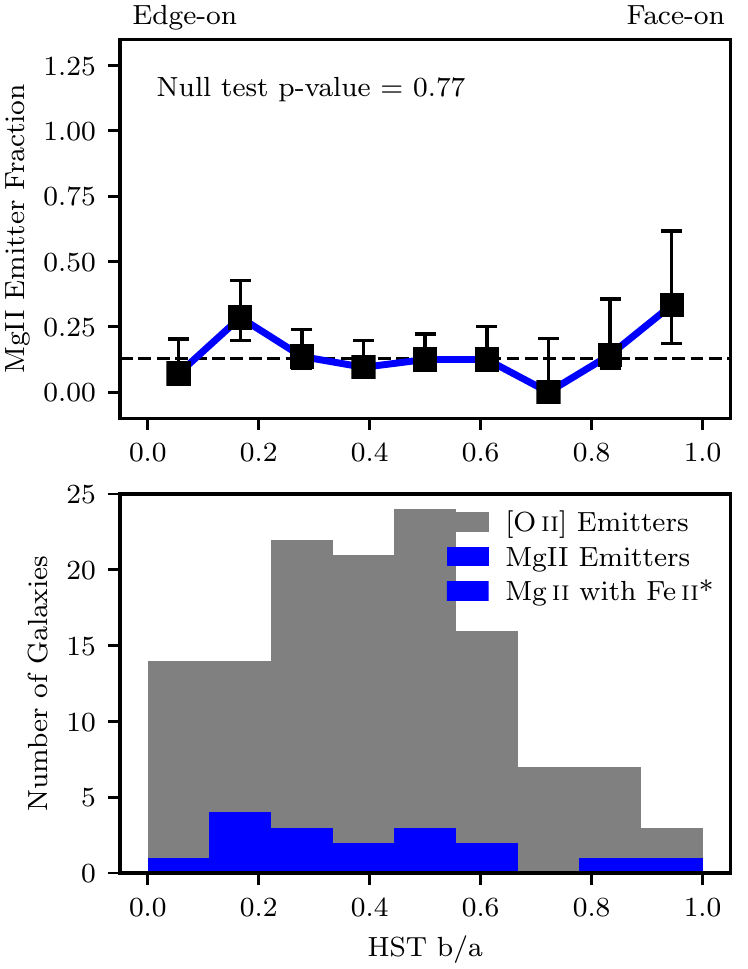}}
\caption{{\bf (a)}: {\it Bottom:} Axis ratio ($b/a$) distribution for the \FeIIs\ emitters from the {\it HST} Y-band. The grey histogram shows the distribution for 69 \OII\ emitters with SFR~$\geq +0.5$~\mpy, and the red histogram shows the subpopulation of \FeIIs\ emitters with confidence flag qc~$> 1$ (23 galaxies). White hatching indicates \FeIIs\ emitters within this SFR range that also have \MgII\ emission or P-Cygni profiles (8 galaxies). {\it Top:} The fraction of \FeIIs\ emitters for the ten axis ratio bins. Error bars represent the 68\%\ confidence interval as in Fig~1.
{\bf (b)}: {\it Bottom:} Axis ratio ($b/a$) distribution for the \MgII\ emitters from the {\it HST} Y-band. The grey histogram shows the distribution for 133 \OII\ emitters with $ -0.5$~\mpy~$\leq $~SFR~$\leq +0.5$~\mpy, and the blue histogram shows the subpopulation of \MgII\ emitters with confidence flag qc~$> 1$ (17 galaxies). White hatching indicates \MgII\ emitters within this SFR range that also have \FeIIs\ emission (1 galaxy). 
{\it Top:} The fraction of \MgII\ emitters for the ten axis ratio bins.
}
\label{fig:morph:ba}
\end{figure*}

We took further advantage of the ancillary data available in the UDF area, and in particular of the size and morphological analysis by \citet{VanderWelA_12a}.  
Briefly, \citet{VanderWelA_12a} performed single Sersic profile fits with the GALFIT \cite{PengC_10a} algorithm on each of the available near-infrared bands (H$_{\rm F160W}$, J$_{\rm F125W}$ and, for a subset, Y$_{\rm F105W}$).
The catalog includes the half-light radius ($R_{\rm eff}$), Sersic index $n$, axis ratio $b/a$, and position angle (PA) for each band. We used the $Y$-band for the analysis of axis ratios and sizes, since it typically has a higher S/N, but found similar results with the other bands.


We explored whether the \FeIIs\ and \MgII\ emitter galaxies have different inclinations or sizes than the \OII\ emitter galaxies for which these signatures are not detected.
To focus on \FeIIs\ emitters, we took only galaxies from the parent sample with log~SFR~$> +0.5$~\mpy, using the SFR values from SED fitting.
This SFR cut includes 69 \OII\ emitters, 23 of which have \FeIIs\ emission with qc~$> 1$. 
Similarly, to focus on \MgII\ emitters, we took only galaxies from the parent sample with $-0.5 \leq {\rm log~SFR} \leq +0.5$~\mpy. This SFR cut includes 133 \OII\ emitters, 17 of which have \MgII\ emission with qc~$> 1$. 
We compare the galaxy properties between \FeIIs\ or \MgII\ emitters and \OII\ emitters within the same SFR range.

Fig.~\ref{fig:morph:ba} shows the axis ratio ($b/a$) distributions for the \FeIIs\ emitters and \MgII\ emitters (bottom panels), as well as the emitter fractions (top panels). 
In both cases, $\chi^2$ statistical tests, as in Sect.~\ref{sec:zdepend}, do not exclude uniform inclination distributions, although the p-value is significantly lower for \FeIIs\ emission (0.09) than for \MgII\ emission (0.96). Nonetheless, neither \FeIIs\ emission nor \MgII\ emission appears to depend on the galaxy inclination.

\begin{figure*}[!t]
\centering
\subfigure[][]{
\includegraphics[width=8cm]{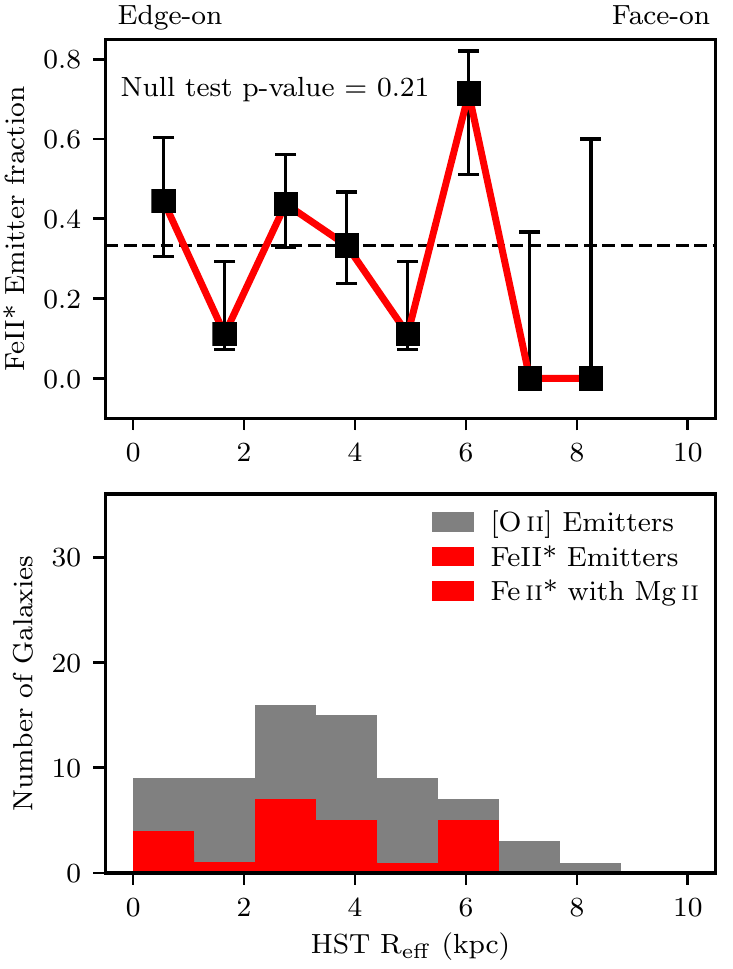}}
\hspace{0.3cm}
\subfigure[][]{
\includegraphics[width=8cm]{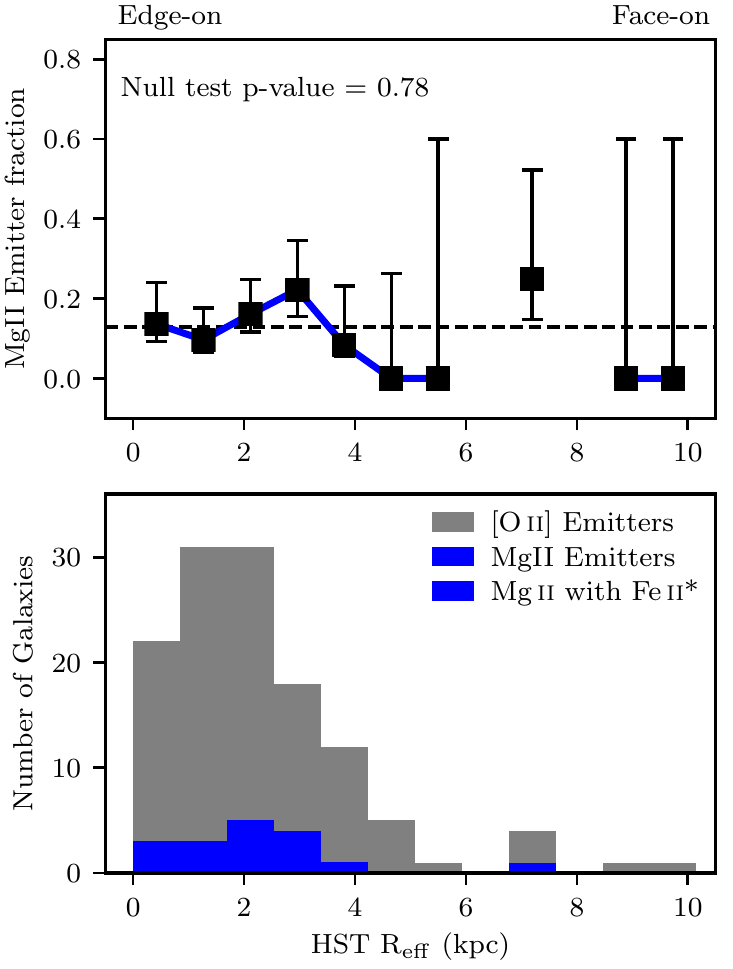}}
\caption{{\bf (a)}: {\it Bottom}: Proper size distribution (R$_{\rm eff}$) for \FeIIs\ emitters based on the {\it HST} Y-band semi-major axis measurements. The grey histogram shows the proper size distribution for 69 \OII\ emitters with SFR~$\geq +0.5$~\mpy. The red histogram shows the subpopulation of \FeIIs\ emitters with confidence flag qc~$> 1$ (23 galaxies). White hatching indicates \FeIIs\ emitters within this SFR range that also have \MgII\ emission or P-Cygni profiles (8 galaxies).
{\it Top}: The fraction of \FeIIs\ emitters. Error bars represent the 68\%\ confidence interval as in Fig~1.
{\bf (b)}: {\it Bottom}: The grey (blue) histogram shows the size distribution for all galaxies (for \MgII\ emitters) respectively.
 {\it Bottom}:
Proper size distribution (R$_{\rm eff}$) for \MgII\ emitters based on the {\it HST} Y-band semi-major axis measurements. The grey histogram shows the proper size distribution for 133 \OII\ emitters with $-0.5$\mpy~$\leq$~SFR~$\leq +0.5$~\mpy. The blue histogram shows the subpopulation of \MgII\ emitters with confidence flag qc~$> 1$ (17 galaxies). White hatching indicates \MgII\ emitters within this SFR range that also have \FeIIs\ emission (1 galaxy).
}
\label{fig:morph:sizes}
\end{figure*}

Similarly, Fig.~\ref{fig:morph:sizes} shows the proper size ($R_{\rm eff}$) distributions for the \FeIIs\ emitters and \MgII\ emitters (bottom panels) and their respective emitter fractions (top panels). 
Applying the $\chi^2$ statistical test to the emitter fractions does not exclude uniform size distributions for the \FeIIs\ and \MgII\ emitters. 
Neither \FeIIs\ emission nor \MgII\ emission appears to depend on the galaxy size.

Having established that \FeIIs\ emitters and \MgII\ emitters do not have inclination or size distributions that are different from their parent populations, we also check whether the \FeIIs\ and \MgII\ distributions are different from each other.
We apply a Kolmogorov–Smirnov (K-S) test to compare the distributions from galaxies with only \FeIIs\ emission and only \MgII\ emission, excluding galaxies that have both emission signatures, which are indicated with white cross hatching in the Fig.s.
The K-S test for the axis ratio distribution does not reject the possibility that the two samples are the same (p-value~$=0.052$), whereas the K-S test for the size distribution (p-value~$=0.033$) does imply that the samples are different. The distribution of \FeIIs\ emitters that do not have accompanying \MgII\ emission peaks at larger sizes than \MgII\ emitter distribution, which is consistent with their higher stellar masses and SFRs.

\subsection{\FeIIs\ and \MgII\ emission as a function of star formation rate surface density}

The star formation rate surface density, $\Sigma_{\rm SFR}$, can be used as a criterion to determine whether a particular galaxy will drive an outflow, since higher SFRs per unit area will produce more pressure to potentially break through the galactic disk. 
The canonical threshold surface density for driving galactic outflows, $\Sigma_{\rm SFR} > 0.1$~\mpy${\rm kpc}^{-2}$, is based on local starburst galaxies \citep{HeckmanT_02a}. However, both recent integral field spectroscopy results from local main sequence galaxies \citep{HoI_2016a} and evidence of galactic outflows within the Milky Way Fermi Bubbles \citep{FoxA_15a, BordoloiR_17a} suggest that galaxies with lower $\Sigma_{\rm SFR}$ values ($\Sigma_{\rm SFR} \approx 10^{-3} - 10^{-1.5}$~\mpy${\rm kpc}^{-2}$) can drive outflows. The threshold surface density may evolve with redshift \citep{SharmaM_16a} and may also depend on the galaxy properties, especially the gas fraction \citep{NewmanS_12a}. The threshold from the $z \sim 2$ \citet{NewmanS_12a} galaxy sample is $\Sigma_{\rm SFR} = 1$~\mpy${\rm kpc}^{-2}$, an order of magnitude above the \citet{HeckmanT_02a} value. Constraints on the threshold surface density will improve as more studies are able characterize both the outflow and the host galaxy properties.

We investigate whether there might be differences in the $\Sigma_{\rm SFR}$ properties for the different populations of emitters. While we previously included P-Cygni profiles in our \MgII\ emitter sample, here we consider galaxies with P-Cygni profiles and pure emission profiles separately.
The pure \MgII\ emitters have a range $-2.6 < {\rm log}~\Sigma_{\rm SFR} < +0.6$~\mpy${\rm kpc}^{-2}$ with mean value $-1.1 \pm 0.7$~\mpy${\rm kpc}^{-2}$.
The \FeIIs\ emitters span a similar range, $-2.7 < {\rm log}~\Sigma_{\rm SFR} < +1.1$~\mpy${\rm kpc}^{-2}$, but with a higher mean value of $-0.6 \pm 0.7$~\mpy${\rm kpc}^{-2}$.
Nearly all of the P-Cygni profile \MgII\ emitters also have \FeIIs\ emission, and they cover the most limited range, $-1.3 < {\rm log}~\Sigma_{\rm SFR} < +0.6$~\mpy${\rm kpc}^{-2}$, with mean value $-0.3 \pm 0.7$~\mpy${\rm kpc}^{-2}$. 
The pure \MgII\ emitters have a lower mean $\Sigma_{\rm SFR}$ value than the \FeIIs\ emitters or the \MgII\ emitters with P-Cygni profiles.

We evaluate whether the pure \MgII\ emitters come from the same distribution as either the \FeIIs\ emitters or the \MgII\ emitters with P-Cygni profiles. In both cases, a K-S test rejects this hypothesis with P-values of 0.02 and 0.01, respectively. 
Pure \MgII\ emitters have a different, lower $\Sigma_{\rm SFR}$ distribution than galaxies with \FeIIs\ emission or \MgII\ P-Cygni profiles, and may be less likely to drive outflows.

\begin{sidewaystable*}
\centering
\small
\caption{Galaxy properties for the \FeIIs\ and \MgII\ emitters in the UDF-10 field, flagged with qc~$> 1$ in the mosaic. }
\begin{tabular}{ccrccllcll}
\hline
GalaxyID & Redshift & \multicolumn{1}{c}{$\log$(\mstar/M$_\sun$)}& $\log$(SFR$_{\rm SED}$/\mpy) & $\log$(SFR$_{\OII}$/\mpy) & \multicolumn{1}{c}{$b/a$} & \multicolumn{1}{c}{$R_{1/2}$} & m$_{F606W}$ & Selection & Comment
\\
 & & \multicolumn{1}{c}{(\msun)} & (\mpy) & (\mpy) &   &  \multicolumn{1}{c}{(kpc)} &   &   &   \\
(1) & (2) & \multicolumn{1}{c}{(3)} & (4) & \multicolumn{1}{c}{(5)} & \multicolumn{1}{c}{(6)} & (7) & (8) & (9)      \\
\hline 
UDF10--0008 & 1.095 & 10.48$^{+0.01}_{-0.21}$ &  2.02$^{0.35}_{0.11}$ & 1.90$\pm0.02$ & 0.80/0.78 & 4.6/5.7 & 22.59 & \FeIIs\ emi & Large face-on, no \MgII\ emission \\
UDF10--0011 & 1.038 & 10.07$^{+0.13}_{-0.11}$ &  1.63$^{0.26}_{0.30}$ & 1.60$\pm0.03$ & 0.38/0.39 & 3.5/5.7 & 23.32 & \FeIIs\ emi & Edge-on, no \MgII\ emission  \\
UDF10--0012 & 0.997 & 10.19$^{+0.05}_{-0.05}$ & -0.24$^{0.05}_{0.01}$ & 1.19$\pm0.03$ & 0.84/0.45 & 1.8/7.0 & 23.98 & \FeIIs\ emi & Edge-on, no \MgII\ emission \\
UDF10--0013 & 0.997 &  9.89$^{+0.17}_{-0.12}$ &  1.06$^{0.28}_{0.21}$ & 1.46$\pm0.12$ & 0.67/0.61 & 2.1/3.5 & 23.52 & \FeIIs\ emi & Face-on, no \MgII\ emission\\
UDF10--0016 & 1.096 & 10.03$^{+0.15}_{-0.14}$ &  1.76$^{0.23}_{0.43}$ & 0.65$\pm0.10$ & 0.71/0.80 & 1.7/2.7 & 24.05 & \FeIIs\ emi & Small face-on, no \MgII\ emission\\
UDF10--0036 & 1.216 & 10.0$^{+0.12}_{-0.44} $ &  0.89$^{0.59}_{0.26}$ & 1.09$\pm0.13$ & 0.93/0.71 & 1.3/1.6 & 25.20 & \FeIIs\ emi & Small compact, no \MgII\ emission \\
UDF10--0032 & 1.307 &  9.23$^{+0.12}_{-0.16}$ &  0.79$^{0.38}_{0.19}$ & 1.44$\pm0.16$ & 0.50/0.33 & 2.2/2.3 & 24.56 & Both & Interaction with 121, edge-on, \MgII\ P-Cygni \\
UDF10--0030 & 1.096 &  8.94$^{+0.18}_{-0.03}$ &  0.97$^{0.09}_{0.32}$ & 1.23$\pm0.12$ &  ~~~--~~/0.54  & ~--~~/0.4  & 24.75 & \MgII\ emi & Compact, weak \FeIIs\ emi, weak \FeII\ abs., \MgII\ P-Cygni\\ 
UDF10--0033 & 1.415 &  9.33$^{+0.44}_{-0.09}$ &  1.06$^{0.21}_{0.42}$ & 1.52$\pm0.14$ & 0.84/0.87 & 2.5/4.3 & 24.61 & \MgII\ emi & Merger compact, weak \FeIIs\ emission, weak \FeII\ abs.\\ 
UDF10--0037 & 0.981 &  8.84$^{+0.13}_{-0.22}$ & -0.19$^{0.67}_{0.11}$ & 0.70$\pm0.12$ & 0.58/0.28 & 1.9/2.8 & 25.17 & \MgII\ emi & No \FeII\ absorption\\
UDF10--0046 & 1.414 &  9.31$^{+0.22}_{-0.20}$ &  0.42$^{0.18}_{0.29}$ & 1.46$\pm0.12$ & 0.32/0.45 & 2.6/2.2 & 25.06 & \MgII\ emi & Merging with 0092, \FeII\ absorption\\
UDF10--0056 & 1.307 &  9.02$^{+0.16}_{-0.13}$ & -0.02$^{0.18}_{0.19}$ & 0.65$\pm0.12$ & 0.72/0.47 & 1.2/1.1 & 25.60 & \MgII\ emi & Small, no \FeII\ absorption\\
UDF10--0092 & 1.414 &  8.54$^{+0.20}_{-0.20}$ & -0.30$^{0.03}_{0.02}$ & 1.13$\pm0.19$ & ~~~--~~/0.18 & ~--~~/1.1 & 26.13 & \MgII\ emi & Small, merging with 0046, maybe \FeII\ absorption \\
\hline
\end{tabular}
\tablefoot{ 
Column (1): Galaxy Name;
Column (2): Redshift;
Column (3): Stellar mass ($\log$ \msun) from SED fitting with the FAST algorithm using a \citet{ChabrierG_03a} IMF.;
Column (4): SFR from SED fitting with the FAST algorithm using a \citet{ChabrierG_03a} IMF.; 
Column (5): SFR from the \OII3727 luminosity (see text) dust corrected using a \citet{ChabrierG_03a} IMF; 
Column (6): Axis-ratio $b/a$  from the \OII\ narrow-band images (intrinsic value, i.e. deconvolved from the seeing) and from {\it HST} Y-band \citep{VanderWelA_12a}; 
Column (7): Half-light radius, $R_{1/2}$ in proper kpc, for the the \OII\ narrow-band images (intrinsic value, i.e. deconvolved from the seeing) and from {\it HST} Y-band \citep{VanderWelA_12a}; 
Column (8): Continuum magnitude from {\it HST} \citep{RafelskiM_15a};
Column (9): Selection according to \FeIIs\ emission (\FeIIs\ emi) or \MgII\ emission (\MgII\ emi);
Column (10): Comment for each galaxy
}
\label{tab:characteristics}
\end{sidewaystable*}

\begin{table*}[ht]
\small
\caption{Rest-Frame Equivalent Width Measurements for the seven \FeIIs\ emitters in the UDF-10 field (Not corrected for emission infilling).}

\begin{tabular}{llcccccccl}
\hline
Multiplet & Line & UDF-0008 & UDF-0011 & UDF-0012 & UDF-0013 & UDF-0016  & UDF-0032 & UDF-0036  \\
(1) & (2) & (3) & (4) & (5) & (6) & (7) & (8) & (9) \\
\hline
UV3 & \FeII${\lambda2344}$      & $+2.11\pm0.19$ & $+1.36\pm0.15$ & --- & --- & $+2.79\pm0.38$ & $+1.56\pm0.23$ & $+1.44\pm0.41$  \\
UV2b & \FeII${\lambda2374}$\tablefootmark{b}    
								& $+1.64\pm0.17$ & $+1.29\pm0.14$ & --- & --- & $+1.81\pm0.36$ & $+1.34\pm0.24$ & $+1.29\pm0.33$ \\
UV2a & \FeII${\lambda2382}$\tablefootmark{a}  
								 & $+2.23\pm0.18$ & $+1.88\pm0.12$ & $-0.01\pm0.33$ & $+0.97\pm0.37$ & $+2.45\pm0.36$ & $+1.35\pm0.24$ & $+1.12\pm0.33$  \\
UV1b & \FeII${\lambda2586}$     	& $+2.24\pm0.15$ & $+2.14\pm0.11$ &  $+2.22\pm0.19$  & $+1.96\pm0.20$ & $+2.31\pm0.30$ & $+1.32\pm0.32$ & $+2.77\pm0.31$ \\
UV1a & \FeII${\lambda2600}$    	& $+2.37\pm0.16$ & $+1.86\pm0.11$ &  $+2.45\pm0.19$  & $+1.68\pm0.21$  & $+3.45\pm0.31$ & $+1.63\pm0.32$ & $+2.20\pm0.33$ \\
--- & \MgII${\lambda2796}$         	&  $+3.58\pm0.15$ & $+2.00\pm0.12$ &  $+1.67\pm0.62$ & $+1.64\pm1.15$  & $+3.71\pm0.29$ & $+0.22\pm0.24$ & $+1.02\pm0.27$ \\
--- & \MgII${\lambda2803}$ 	     	&  $+3.23\pm0.16$ & $+1.91\pm0.12$ &  $+0.86\pm0.14$ & $+2.18\pm0.22$ & $+4.06\pm0.33$ & $+0.70\pm0.31$ & $+1.31\pm0.26$\\
 --- & \MgI${\lambda2852}$        	&  $+0.83\pm0.15$ & $+0.66\pm0.11$  &  $+1.22\pm0.16$ & $+0.74\pm0.21$ & $+0.81\pm0.34$ & $+1.10\pm0.28$ & $+0.69\pm0.26$ \\
UV3 & \FeIIs${\lambda2365}$    & $-0.05\pm0.13$ & $-0.36\pm0.10$ & --- & --- & $-0.04\pm0.22$ &  $-0.43\pm0.21$ & $-0.54\pm0.28$ \\
UV3 & \FeIIs${\lambda2381}$\tablefootmark{a}  
								&  --- &  --- &  ---  &  ---  &  --- &  --- &  ---  \\
UV2b & \FeIIs${\lambda2396}$\tablefootmark{c}  
								 & $-0.11\pm0.12$ & $-0.70\pm0.10$ & $-0.99\pm0.23$ & $-0.62\pm0.25$  & $-0.57\pm0.25$ & $-0.55\pm0.21$ & $-0.99\pm0.27$  \\

UV1a & \FeIIs${\lambda2612}$  	& $-0.46\pm0.11$ & $-0.23\pm0.08$ & $-1.21\pm0.18$ & $-0.47\pm0.21$ & $-0.06\pm0.20$ & $-0.44\pm0.22$ & $-0.88\pm0.29$ \\
UV1a & \FeIIs${\lambda2632}$ 	& --- & --- & --- & --- & --- & --- & ---   \\
UV1b & \FeIIs${\lambda2626}$   	& $-0.81\pm0.11$ & $-0.26\pm0.08$ & $-2.12\pm0.17$ & $-0.82\pm0.21$ & $-0.33\pm0.20$  & $-0.79\pm0.22$ & $-1.09\pm0.25$ \\
--- & \OII${\lambda3727}$       & $-21.1\pm0.3$~~ & $-34.1\pm0.2$~~  & $-39.6\pm0.1$~~ & $-42.0\pm0.2$~~   &  $-20.2\pm0.6$~~ & $-69.7\pm0.6$~~ & $-56.1\pm0.3$~~ \\

\hline 
\end{tabular}
\tablefoot{Column (1): Multiplet;
Column (2): Transition wavelength;
Column (3)--Column (7): rest-frame equivalent width for each galaxy. Emission is negative and absorptions is positive. 
}
\tablefoottext{a}{\FeII$\lambda$2382 is a pure resonant absorption line with no associated \FeIIs\ emission, but it is blended with the weak \FeIIs$\lambda$2381 emission from the UV3 multiplet.}\\
\tablefoottext{b}{\FeII$\lambda$2374 is effectively free of emission infilling, because nearly all photons absorbed at \FeII$\lambda$2374 are re-emitted at the non-resonant \FeIIs$\lambda2396$ line \citep{TangY_14a,ZhuG_15a}.}\\
\tablefoottext{c}{\FeIIs$\lambda$2396 is therefore an almost purely fluorescent emission line, since $\sim 90$\% of photons absorbed at \FeII$\lambda$2374 are re-emitted at the non-resonant \FeIIs$\lambda2396$ line \citep{TangY_14a,ZhuG_15a}.}
\label{tab:measures}
\end{table*}

\section{Representative Cases}
\label{sect:representative}

In  section~\ref{sec:mainseq}, we observed a dichotomy along the main sequence between galaxies with only \MgII\ emission and galaxies with only \FeIIs\ emission. Furthermore, these emitters appear to show a progression where galaxies with \mstar~$\lesssim 10^{9}$~\msun\ tend to have only \MgII\ emission with no accompanying \MgII\ or \FeII\ absorption features, galaxies at the transition around \mstar~$\sim 10^{9.5}$~\msun have \MgII\ P-Cygni profiles with moderate \FeII\ absorption with \FeIIs\ emission, and galaxies with \mstar~$\gtrsim 10^{10}$~\msun\ have strong \MgII\ and \FeII\ absorption profiles with \FeIIs\ emission.

In order to investigate the 1D spectral properties of a representative sample, we selected galaxies that are detected in the deeper UDF-10 field in order to benefit from the higher signal-to-noise.
Of the 25 \FeIIs\ emitters with qc~$ > 1$ in our UDF mosaic sample, seven are in the UDF-10 field, one of which is also detected with \MgII\ emission. 
Of the 33 \MgII\ emitters with qc~$ > 1$ in the mosaic, seven are in UDF-10 field. 
Two of these \MgII\ emitters have P-Cygni profiles. 
We summarize the characteristics of the 13 UDF-10 galaxies in Table~\ref{tab:characteristics}.

\begin{figure*}
\centering
\includegraphics[]{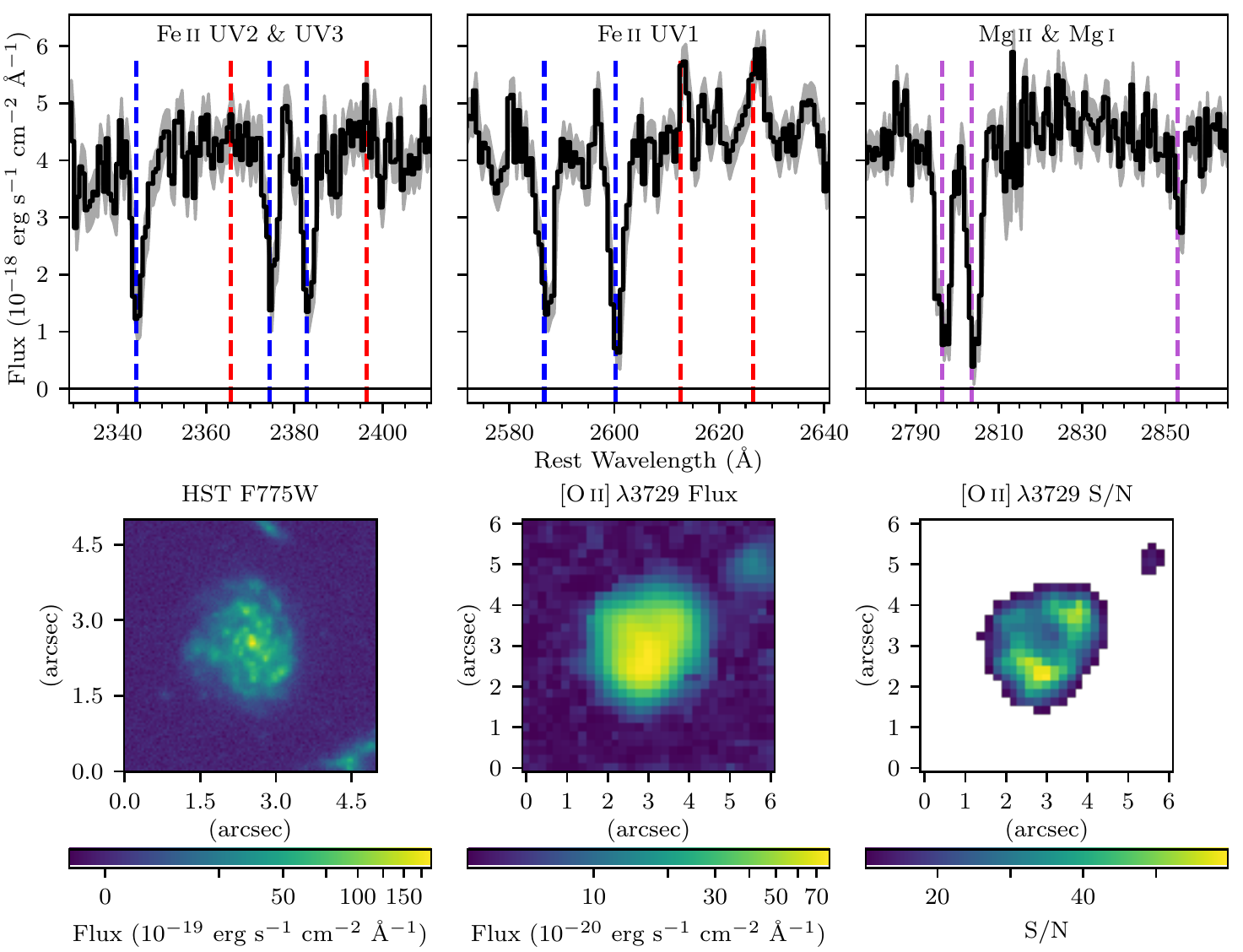}
\caption{UDF Galaxy ID~8 at $z = 1.0948$. The top row shows sections of the MUSE spectrum with the UV2 and UV3 \FeII\ multiplets (\FeII\,$\lambda2344$, \FeIIs$\lambda2365$, \FeII\,$\lambda\lambda2374,2382$ and \FeIIs$\lambda2396$), the UV1 \FeII\ multiplet (\FeII\,$\lambda\lambda2586,2600$ and \FeIIs$\lambda\lambda2612,2626$), and \MgII\,$\lambda\lambda2796,2803$ with \ion{Mg}{I}\,$\lambda2852$.
The blue (purple) dashed lines indicate the resonant \FeII\ (\MgII) transitions, and the red dashed lines show the non-resonant \FeIIs\ emission.
The bottom row shows the $HST$ F775W image and the MUSE the \OII\,$\lambda3729$ flux map with an asinh scale, along with the corresponding MUSE S/N map with a threshold of S/N~$> 10$. 
This galaxy is large and face-on. The spectrum shows \FeII, \MgII, and \MgI\ absorption features, with \FeIIs\ emission. 
}

\label{fig:udf10-id08}
\end{figure*}

\begin{figure*}
\centering
\includegraphics[]{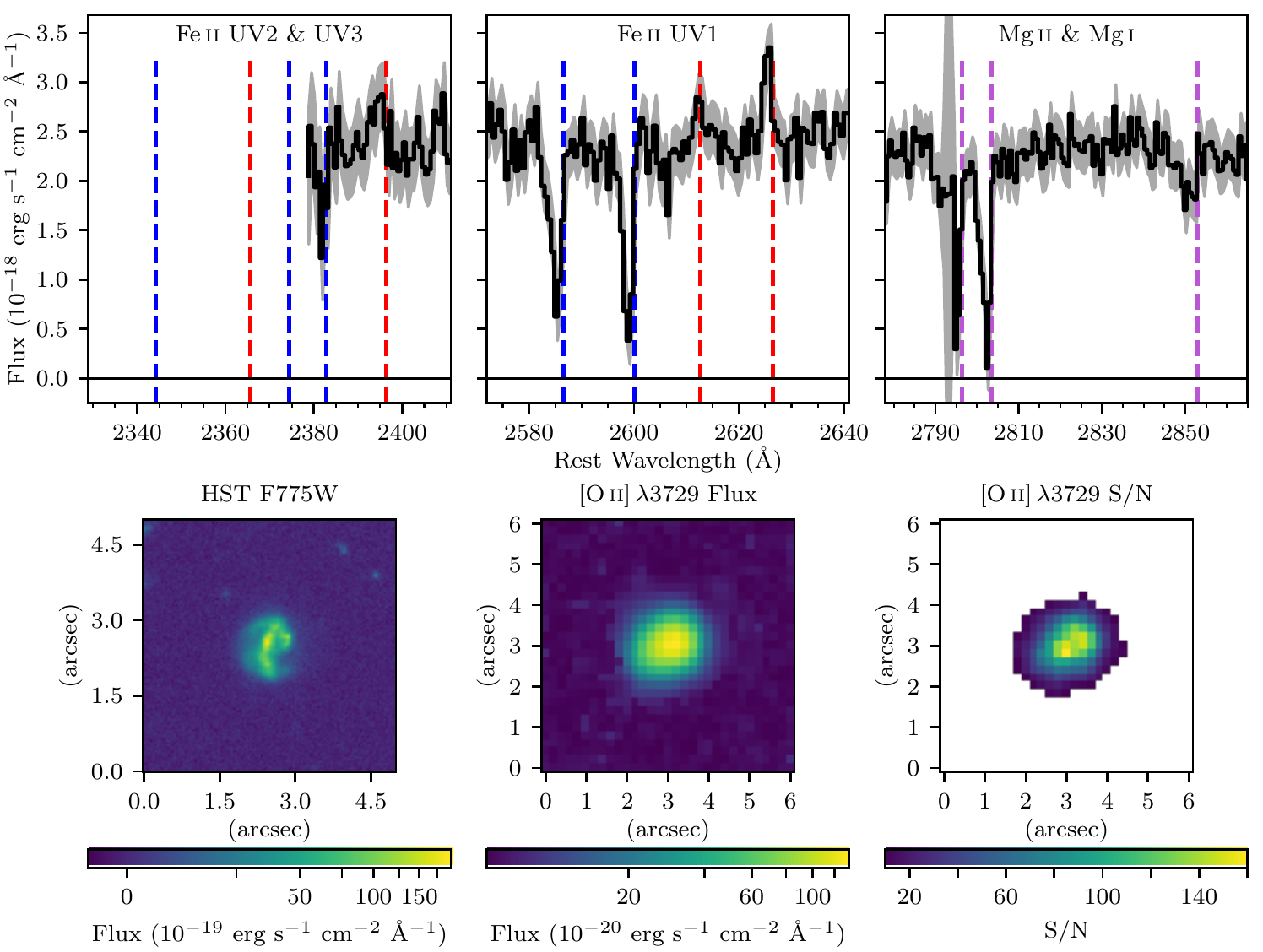}
\caption{UDF Galaxy ID~13 at $z = 0.9973$. Same panels as Fig.~\ref{fig:udf10-id08}. For this redshift, the \FeII\ UV2 \& UV3 multiplets are not fully covered in the MUSE spectral range. 
Like the galaxy ID~8 (Fig.~\ref{fig:udf10-id08}), this galaxy appears to be face on but disturbed, and the spectrum shows \FeII, \MgII, and \MgI\ absorption features, with \FeIIs\ emission. }
\label{fig:udf10-id13}
\end{figure*}

\begin{figure*}
\centering
\includegraphics[]{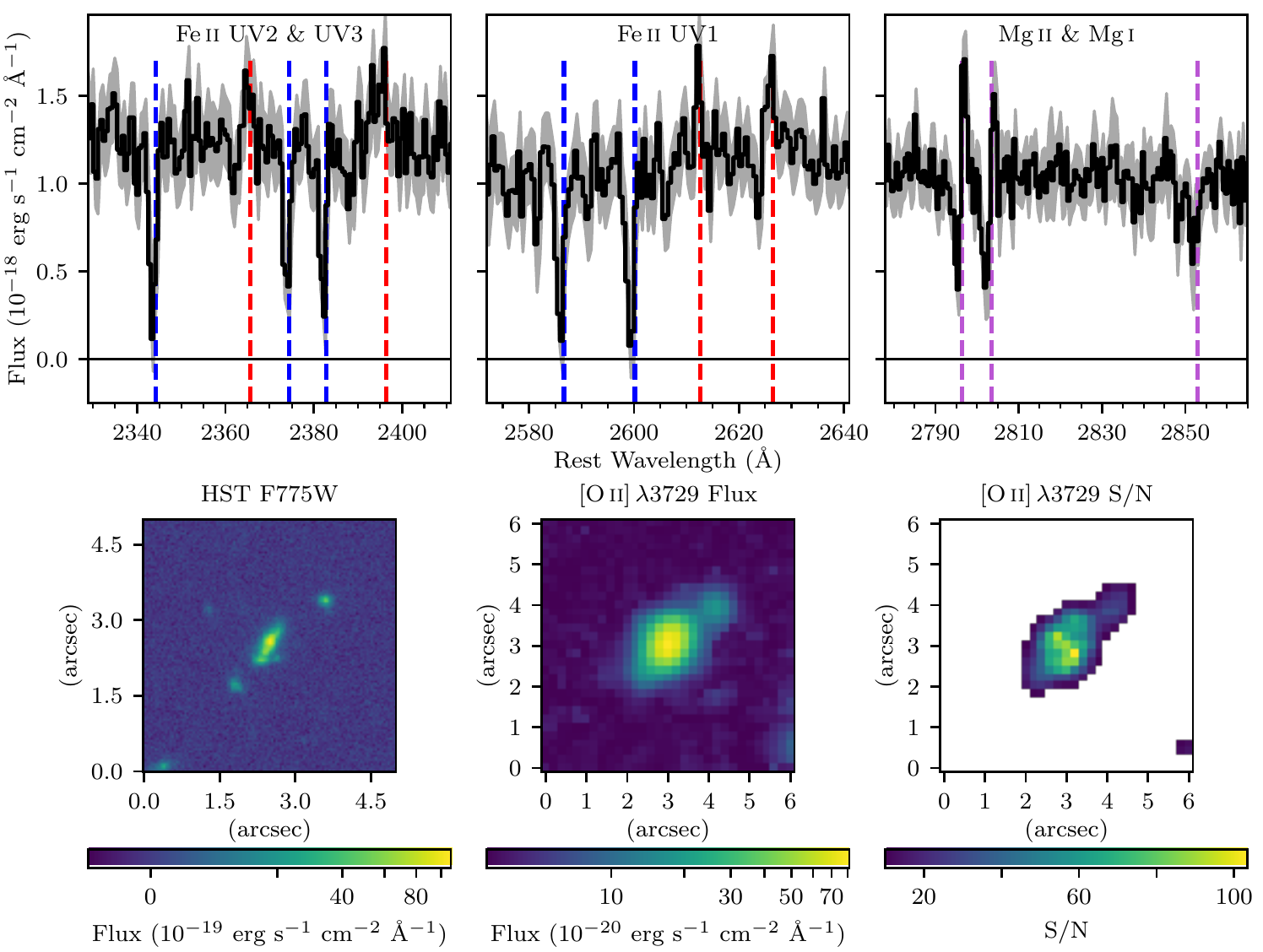}
\caption{UDF Galaxy ID~32 at $z = 1.3071$. Same panels as Fig.~\ref{fig:udf10-id08}. 
This galaxy appears to be edge-on and is merging with UDF Galaxy ID~121. The spectrum shows \FeII, \MgII, and \MgI\ absorption features, with \FeIIs\ emission and a P-Cygni profile for \MgII. }
\label{fig:udf10-id32}
\end{figure*}

\begin{figure*}
\centering
\includegraphics[]{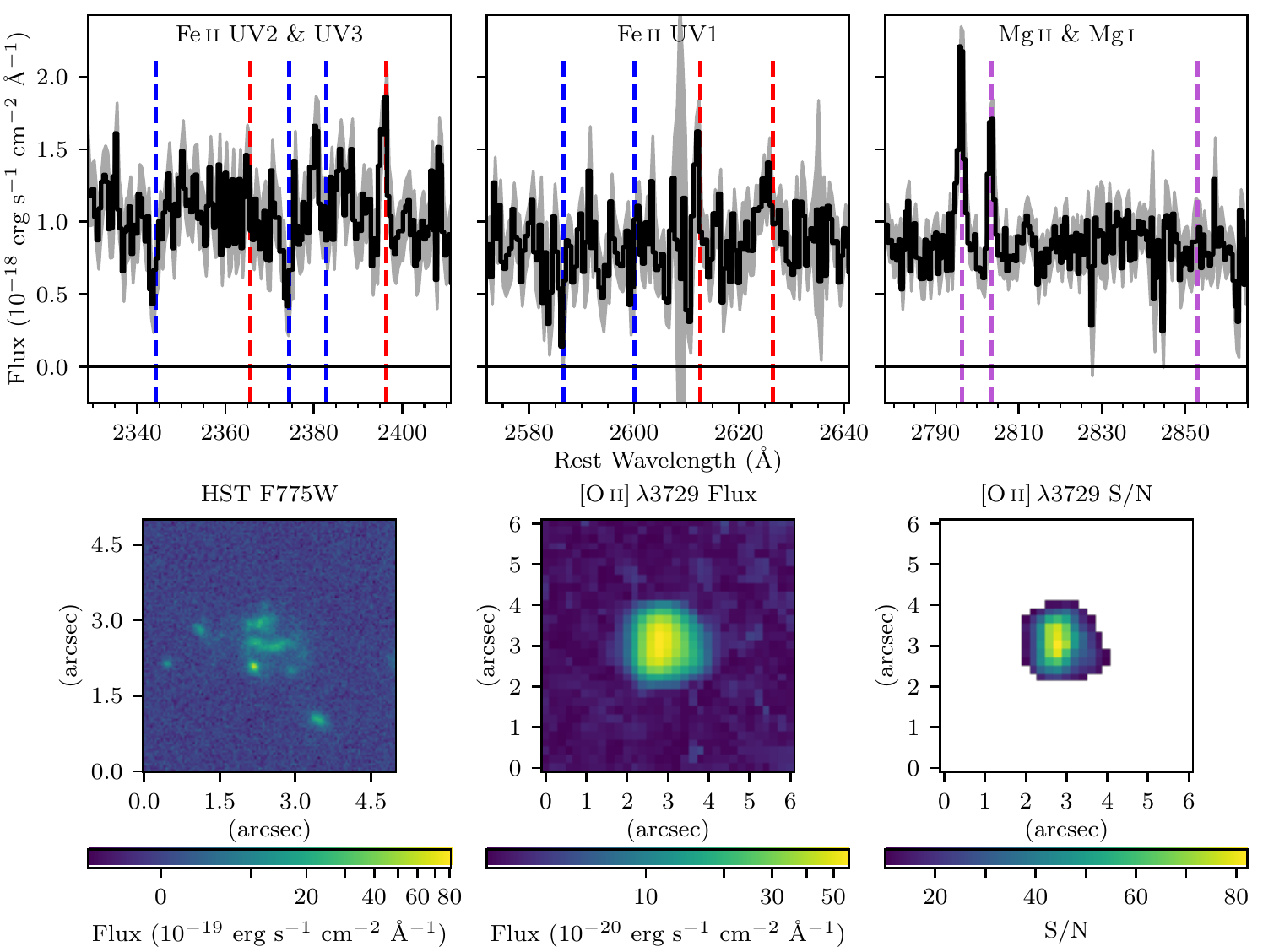}
\caption{UDF Galaxy ID~33 at $z = 1.4156$. Same panels as Fig.~\ref{fig:udf10-id08}. 
Based on the HST image, this galaxy appears to be merging. The spectrum shows weak \FeII\ absorption (most apparent for \FeII\,$\lambda$2344 and \FeII\,$\lambda$2374), weak \FeIIs\ emission, and strong \MgII\ emission. }
\label{fig:udf10-id33}
\end{figure*}

\begin{figure*}
\centering
\includegraphics[]{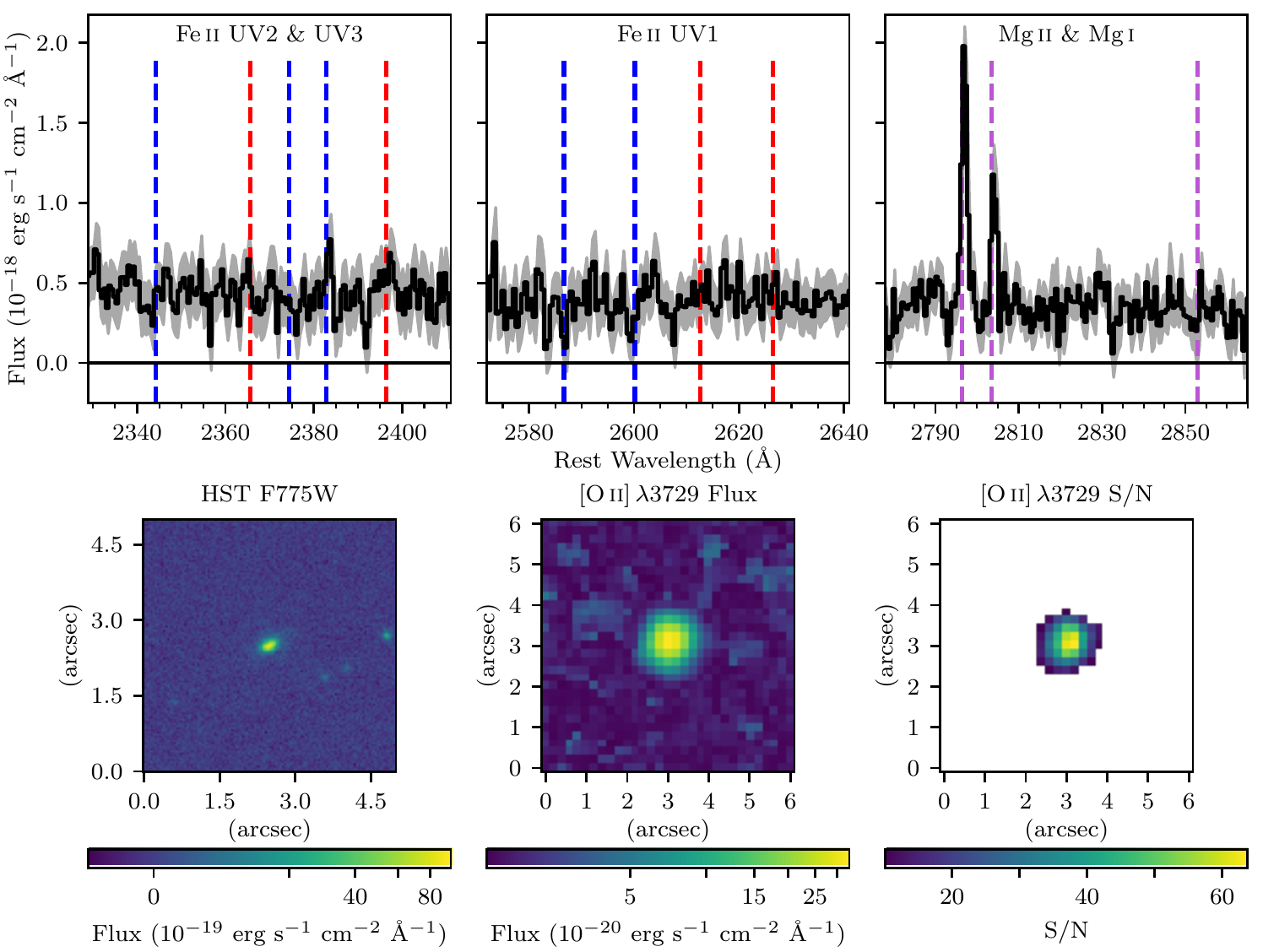}
\caption{UDF Galaxy ID~56 at $z = 1.3061$. Same panels as Fig.~\ref{fig:udf10-id08}. 
This galaxy is compact, and the spectrum shows only \MgII\ emission, without \FeII\ absorption or \FeIIs\ emission. The \MgII\ absorption creating a slight P-Cygni profile for this \MgII\ emitter is detectable only in the UDF-10 spectrum.}
\label{fig:udf10-id56}
\end{figure*}

Fig.s~\ref{fig:udf10-id08}--\ref{fig:udf10-id56} transition from examples of galaxies with strong \MgII\ absorption (ID08 and ID13) to a P-Cygni profile (ID 32) to strong \MgII\ emission (ID 33 and ID 56). 
All of these galaxies, except for ID 56, also have \FeIIs\ emission and \FeII\ absorption. However, the weak \FeIIs\ emission and \FeII\ absorption for ID33 are detected only in the UDF-10 spectrum, not flagged in the mosaic. 
The \FeIIs\ emitters flagged from the mosaic (Fig.s~\ref{fig:udf10-id08}--\ref{fig:udf10-id32}) all have \FeII\ and \MgII\ in absorption, with possible emission infilling (see next section). 
Interestingly, the \MgII\ emitters are often associated with a merging event, such as ID33, ID46 with ID92, and ID32 with ID121.
Merging events may provoke outflows from these lower mass galaxies. The P-Cygni profile from ID33 is further evidence of an outflow.

\subsection{Emission Signature Properties  from 1D Spectra}

For each of the seven \FeIIs\ emitters in the UDF-10 field, we measured the rest-frame equivalent widths for the \FeII\ absorption and \FeIIs\ emission  (Table~\ref{tab:measures}) from the PSF-weighted sky-subtracted spectrum. 
For each spectrum, we fit the continuum with a cubic spline using a custom interactive python tool. From the normalized spectrum, we measure the rest-frame equivalent widths over velocity ranges that cover the full absorption/emission profiles. We calculate the equivalent widths by directly summing the flux and estimate uncertainties on these equivalent widths from the noise of the spectrum. 

Before quantifying the the equivalent widths, we note that
 \FeII\ and \MgII\ absorption lines may be affected by emission infilling \citep{ProchaskaJ_11a,ScarlataC_15a, ZhuG_15a}. Emission infilling occurs when an absorbed photon is re-emitted at the same wavelength, producing underlying emission that fills in the absorption profile and can shift the maximum absorption profile depth blueward. At its most extreme, emission infilling produces P-Cygni profiles. 
Emission infilling affects some transitions more than others, depending on how likely it is for the absorbed photon to be re-emitted resonantly.
From \citet{ZhuG_15a}, the probability of emission infilling for each of the resonant \FeII\ transitions is:
\begin{equation}
p_{\FeII}^{\lambda2374} < 
p_{\FeII}^{\lambda2586} < 
p_{\FeII}^{\lambda2344} < 
p_{\FeII}^{\lambda2600} <
p_{\rm res} ,
\end{equation}
where $p_{\rm res}$ is the probability of emission infilling for purely resonant transitions that do not have associated non-resonant transitions, such as \FeII\,$\lambda2383$ and \MgII.
For purely resonant transitions, the amount of emission infilling depends mainly on the degree of saturation, which in turn follows the absorption strength.
Based on the elemental abundance and oscillator strength for each transition, the expected order for the absorption strength from \citet{ZhuG_15a} is:
\begin{eqnarray}
W_{\MgI}^{\lambda2852} < 
W_{\FeII}^{\lambda2383} < 
W_{\MgII}^{\lambda2803} <
W_{\MgII}^{\lambda2796}  .
 \end{eqnarray}
The \MgII\ doublet is therefore the most susceptible to emission infilling. Among the \FeII\ transitions, \FeII$\lambda2383$ is the most susceptible, while \FeII\,$\lambda2374$ and $\lambda2586$ are the least susceptible to emission infilling.
The radiative transfer models from \citet{ProchaskaJ_11a} and \citet{ScarlataC_15a} have shown that the amount of observed emission infilling also depends on several other factors, such as the outflow geometry and dust content.
  
\begin{figure}[!ht]
\centering
\includegraphics[width=7.5cm]{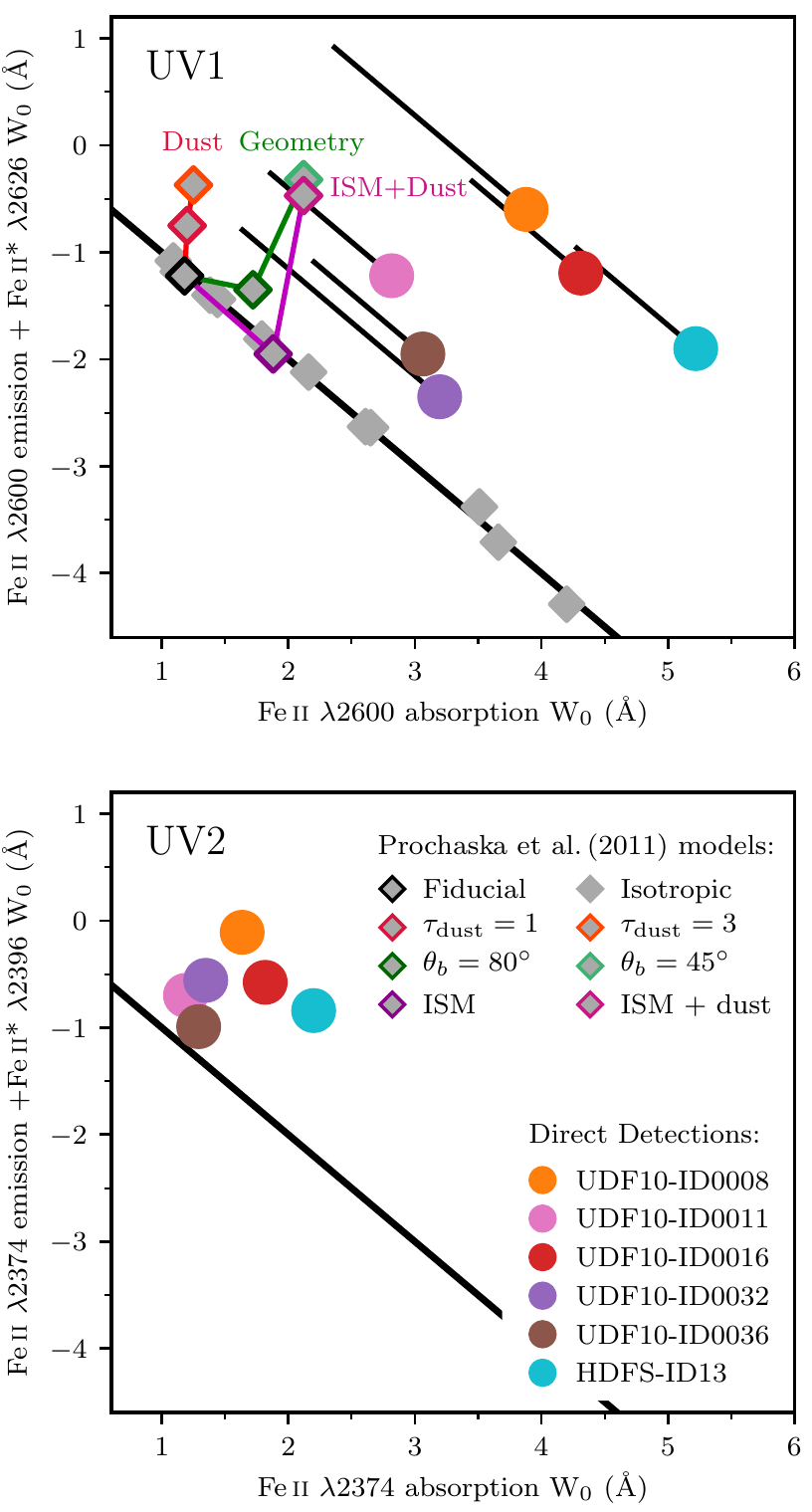}
\caption{Photon-conservation diagnostics for the two resonant \FeII\ transitions, $\lambda2374$ UV2 (bottom) and $\lambda2600$ UV1 (top), with only one \FeIIs\ re-emission channel (\FeIIs$\lambda2396$ and \FeIIs$\lambda2626$ respectively). In both panels, the $x$-axis is the resonant absorption equivalent width, and the $y$-axis is the total re-emission equivalent width from the resonant and non-resonant transitions. However, resonant re-emission (emission infilling) is negligible for the \FeII$\lambda2374$ transition \citep{TangY_14a,ZhuG_15a}. The diagonal black line represents photon-conservation between emission and absorption processes.  
The solid colored points represent the \FeIIs\ emitters from this sample with the emission infilling correction (see text).
The black lines associated with these points in the top panel trace the difference between the measured and corrected equivalent width values. 
The solid diamonds represent theoretical predictions from the radiative transfer models of \citet{ProchaskaJ_11a}. Gray diamonds indicate isotropic outflow models, which all respect photon conservation, and the diamonds with colored outlines show variations to the geometry and dust content that decrease the total amount of re-emission.
}
\label{fig:diagnostic}
\end{figure}

We now quantify the amount of infilling for the \FeIIs\ emitters from the rest-frame equivalent widths measurements using the \citet{ZhuG_15a} method. This method consists of comparing the observed rest-frame equivalent widths of the resonant lines detected in galaxy spectra to those seen as intervening absorption systems in quasar spectra (see their Fig. 12). The \FeII\,$\lambda2374$ transition is the anchor point for this correction, since it is the least affected by emission infilling, as discussed in \citet{TangY_14a} and \citet{ZhuG_15a}. Here, we take the averaged rest-frame equivalent widths of resonant \FeII\ and \MgII\ absorption from a stacked spectrum of $\sim 30$ strong \MgII\ absorber galaxies at $0.5<z<1.5$ from \citet[][their Table 7]{DuttaR_17a} as a reference for intervening systems. The top panel of Fig.~\ref{fig:diagnostic} shows the impact of the correction with diagonal black lines that trace the changes to the equivalent width values measured from each galaxy.

In Fig.~\ref{fig:diagnostic}, we follow \citet{ErbD_12a} and compare the amount of absorption on the $x$-axis with the total amount of emission (resonant and non-resonant) on the $y$-axis for the UV1 \FeII\,$\lambda2600$ (top) and UV2 \FeII\,$\lambda2374$ (bottom) transitions.  Of the UV1, UV2, and UV3 \FeII\ multiplets, these are the only transitions that have a single \FeIIs\ re-emission channel. For the UV2 \FeII\,$\lambda2374$ transition (bottom), $\sim90$\%\ of the re-emission is through the non-resonant channel, \FeIIs$\lambda2396$, such that the resonant emission can be neglected. 
Resonant re-emission impacts the \FeII\,$\lambda2600$ transitions more significantly, since only 13\% of the re-emission is through the non-resonant \FeIIs\,$\lambda2626$ transition in a single-scattering approximation \citep{TangY_14a}.
The blue solid line represents the case of photon-conservation, where all of the absorbed photons are re-observed as resonant and non-resonant emission.

The solid colored points in Fig.~\ref{fig:diagnostic} indicate the \FeIIs\ emitter equivalent widths for the UDF-10 sub-sample, along with the HDFS-ID13 $z=1.29$ galaxy from \citet{FinleyH_17a}. Here, the observed resonant \FeII\ absorption and emission equivalent widths (Table~\ref{tab:measures}) are corrected using the infilling emission correction for the UV1 \FeII\,$\lambda2600$ transition as discussed earlier. The solid black lines trace the difference between the measured and the corrected values.
This infilling correction moves points parallel to the photon-conservation line, since accounting for emission infilling increases both the amount of absorption and the total amount of emission.
The galaxies that are furthest from the photon conservation line are all larger face-on galaxies, characteristics that facilitate detecting absorption.

The diamonds in Fig.~\ref{fig:diagnostic} represent theoretical predictions for the UV1 \FeII\,$\lambda2600$ and \FeIIs$\lambda2626$ transitions from the \citet{ProchaskaJ_11a} radiative transfer models of galactic outflows. No models are available for the UV2 \FeIIs$\lambda2374$ transition. 
The fiducial model (black outlined diamond) assumes a dust-free, isotropic radial outflow with the gas density decreasing as $r^{-2}$ and the velocity decreasing as $r$. 
Variations on the fiducial model test additional gas density and velocity laws (gray diamonds), and these models, like the fiducial model, follow the photon-conservation line.
Some of the isotropic, dust-free models predict \FeII\,$\lambda2600$ absorption values of $W_0 \sim 3 - 4 ~\AA$, similar to what is observed for the \FeIIs\ emitter galaxies. However, they all over-predict the corresponding total amount of emission.

The diamonds with colored outlines in Fig.~\ref{fig:diagnostic} show models that deviate from the photon-conservation line and predict more absorption than emission. These models test the effects of dust extinction or collimated outflow geometries. 
Increasing the dust extinction in an isotropic outflow model (red and orange outlined diamonds) decreases the total amount of re-emission and produces a nearly vertical offset from the photon-conservation line. 
The impact of dust extinction becomes more pronounced after introducing a component that represents the interstellar medium (ISM), i.e., gas that is centralized and lacks a significant radial velocity.
Adding only the ISM component shifts the model predictions along the photon-conservation line (purple outlined diamond), whereas including an ISM component plus $\tau_{\rm dust} = 1$ dust extinction (magenta outlined diamond) significantly decreases the total amount of re-emission.

Finally, modifying the outflow geometry such that it becomes increasingly collimated ($\theta_b = 80^{\circ}, 45^{\circ}$, green outlined diamonds) also moves the model predictions away from the photon-conservation line.
Interestingly, the highly collimated outflow model (light green outlined diamond) and the isotropic outflow with an ISM component and dust extinction (magenta outlined diamond) both occupy the same parameter space in this Fig., despite having very different physical properties. Additional modeling is required to better understand the combined effects of dust extinction and geometry.

Comparing the top and bottom panels of Fig.~\ref{fig:diagnostic} shows that, irrespective of the infilling correction, the observed data for the \FeII\,$\lambda2600$ transition is more offset from the photon-conservation line than the \FeII\,$\lambda2374$ transition from the same galaxy. 
Dust extinction can account for both the offset from the photon-conservation line as well as why this offset is more pronounced for the \FeII\,$\lambda2600$ transition. 
The \FeII\,$\lambda2600$ transition is more sensitive to dust extinction, since this transition is more likely to produce resonant re-emission than \FeIIs\,$\lambda2626$ non-resonant emission following a single scattering process \citep[see ][their Fig. 5]{TangY_14a}.
The resonant re-emission undergoes multiple scatterings, and photons that repeatedly scatter also have multiple chances to be absorbed by dust, a process known as resonant trapping.
Conversely, the \FeII$\lambda2374$ transition is less sensitive to dust extinction and resonant trapping, since nearly all of the re-emission is through the non-resonant \FeIIs$\lambda2396$ channel.
Thus, for a given galaxy, the \FeII\,$\lambda2600$ transition has a larger offset from the photon-conservation line than the \FeII$\lambda2374$ transition, due to its greater sensitivity to dust extinction.

\begin{figure}[!ht]
\centering
\includegraphics[width=7.5cm,trim=0 7.8cm 0 0,clip=True]{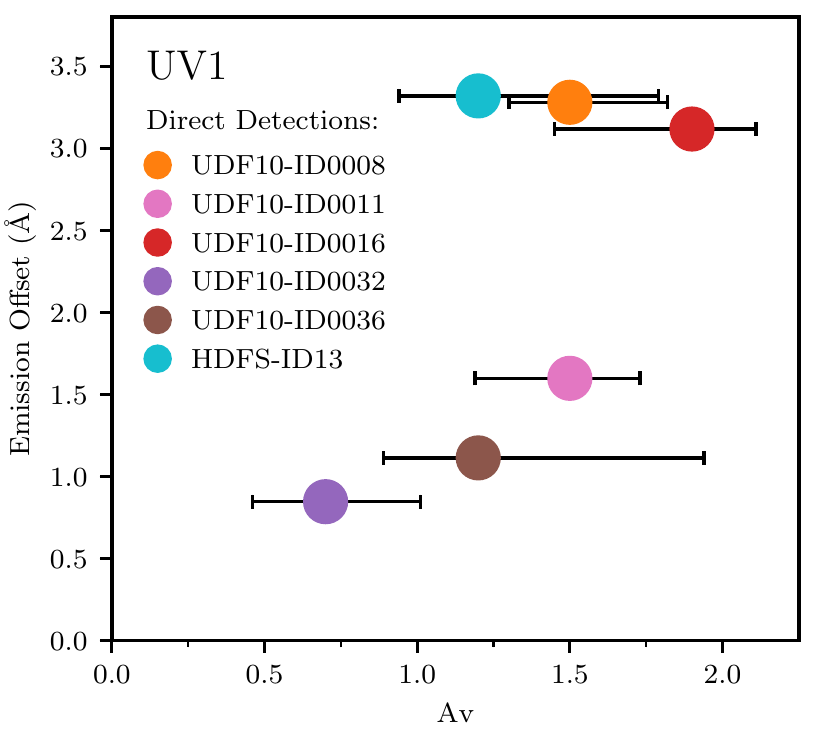}
\caption{Emission offset versus dust extinction from SED fitting.
For each galaxy from Fig.~\ref{fig:diagnostic}, the emission offset measures the vertical distance between the total emission from \FeII\,$\lambda2600$ and \FeIIs$\lambda2626$ in the UV1 multiplet and the photon-conservation line.
}
\label{fig:offsetAV}
\end{figure}

Fig.~\ref{fig:offsetAV} quantifies the vertical offset between the total \FeII\,$\lambda2600$ and  \FeIIs\,$\lambda2626$ re-emission from each galaxy and the photon-conservation line and suggests that the emission offset might increase with increasing dust extinction.
The emission offset and dust extinction in Fig.~\ref{fig:offsetAV} have a Pearson correlation coefficient of 0.63, but more data points are necessary to solidify the trend.
The dust extinction estimate, A$_V$, is from SED modeling (Sect.~\ref{sec:mainseq}), which is robust for the UDF-10 galaxies given the deep $HST$ imaging across multiple bands.
Dust extinction is potentially a significant factor contributing to the offset between the observed emission and the photon-conservation line, in agreement with the \citet{ProchaskaJ_11a} radiative transfer models.
The other significant factor driving the offset may be geometric effects, as discussed above. 
However, more models are required to determine how to best characterize the impact of geometric effects and compare this impact with that of dust extinction.

\section{Discussion}
\label{discussion}

Along the SFR main sequence (Fig.~\ref{fig:Fest:MS}), the emission signatures vary from only \MgII\ emission, to both \MgII\ and \FeIIs\ emission, to only \FeIIs\ emission. 
We propose that this progression is physically motivated, with distinct physical processes producing the emission signatures at the two extremes of the SFR main sequence.

The physical processes that produce \FeIIs\ emission at the high mass, high SFR end and \MgII\ emission at the low mass, low SFR end may be distinct. 
For \FeIIs, the physical process driving non-resonant emission is continuum fluorescence \citep{ProchaskaJ_11a}\footnote{While it is also possible to produce \FeIIs\ emission through indirect UV pumping or collisional excitation, indirect UV pumping requires close proximity ($< 100$~pc) to strong UV sources, and collisional excitation requires high density environments with $> 10^5$~cm$^{-2}$.}. 
For \MgII, two main physical processes can give rise to emission in low mass galaxies: resonant scattering following continuum absorption or nebular emission in \ion{H}{ii} regions\footnote{AGN or shocks from merger events can also produce \MgII\ emission.}.
Whether \MgII\ emission in a particular galaxy is predominantly due to continuum scattering or nebular emission may depend on the strength of the stellar continuum, which we quantify with the \textit{HST F606W} magnitude.

The low mass, low SFR galaxies with only \MgII\ emission detected have weak stellar continua (m$_{F606W} \approx 26$).
When fewer continuum photons are available to undergo absorption, less continuum scattering and less \FeIIs\ emission occurs. 
Galaxies with weak stellar continua therefore do not have significant absorption or \FeIIs\ emission features; instead, they have \MgII\ emission as the predominant feature. 
The \MgII\ emission has median equivalent width values of W$_{0, 2796} = -4.1$~\AA\ and W$_{0, 2803} = -1.7$~\AA, with a typical error (median $1\sigma$ measurement error) of $1.3$~\AA.
\MgII\ emission alone, without accompanying \FeIIs\ emission, likely comes predominantly from \ion{H}{ii} regions, rather than from continuum scattering. 
Based on photoionization modeling, most \MgII\ emission in $1 < z < 2$ star-forming galaxies with outflow signatures is from \ion{H}{ii} regions, but these \ion{H}{ii} regions would need higher ionization parameters to directly produce the \FeIIs\ emission \citep{ErbD_12a}.

As the strength of the stellar continuum increases (m$_{F606W} \approx 24.6$), 
the galaxy spectra show \FeII\ absorption and \FeIIs\ emission, along with \MgII\ P-Cygni profiles. 
The \MgII\ P-Cygni profiles, which are overall dominated by absorption, have median equivalent width values of W$_{0, 2796} = +0.7$~\AA\ and W$_{0, 2803} = +1.1$~\AA, with a typical error of $0.5$~\AA.
The appearance of \MgII\ P-Cygni profiles suggests that continuum scattering is the physical process driving \MgII\ emission in these galaxies.
Previously studied direct detections of \FeIIs\ emission and \MgII\ P-Cygni profiles in individual galaxies \citep{RubinK_11a,MartinC_13a} both demonstrate that continuum scattering in galactic outflows produce these emission signatures.

Finally, the high mass, high SFR galaxies with \FeIIs\ emission but no accompanying \MgII\ emission have the strongest stellar continua of the sample (m$_{F606W} \approx 23.6$) and strong \FeII\ and \MgII\ absorption features.
The \MgII\ absorptions have median equivalent width values of W$_{0, 2796} = +2.7$~\AA\ and W$_{0, 2803} = +2.2$~\AA, with a typical error of $0.3$~\AA.
In the case of strong absorption, the absorbed continuum photons can become resonantly trapped, i.e., they undergo so many scattering events that few photons escape as resonant emission. 
Resonant trapping suppresses emission from the \MgII\,$\lambda\lambda2796,2803$ transitions, which are purely resonant with no non-resonant channels.
However, resonant trapping promotes \FeIIs\ emission, since more scattering events provide more opportunities for photons to escape through a non-resonant channel.
Due to resonant trapping, stronger absorption features imply weaker \MgII\ emission.

Since dust extinction enhances resonant trapping, we can expect to see more \MgII\ emission from galaxies with less dust.  
Dust extinction increases with the galaxy mass and SFR \citep[e.g.,][]{KewleyL_04a, BrinchmannJ_04a}, so the low-mass, low-SFR \MgII\ emitters likely have the least amount of dust. 
Indeed, Feltre et al. (in preparation) find lower extinction values for the MUSE UDF \MgII\ emitters compared to \MgII\ absorbers.
Similarly, dust extinction is potentially the driving factor that determines the strength of \FeIIs\ emission \citep{KorneiK_13a}. 
While resonant trapping from strong absorption components enhances the \FeIIs\ emission, dust extinction from these same components mitigates this enhancement. 
We can expect a trend between the dust extinction and the amount of re-emission (explored in Fig.~\ref{fig:offsetAV}), which may become clearer if we considered only the ISM component.

The physical process driving the \MgII\ and \FeIIs\ emission signatures helps determine whether these signatures trace galactic outflows. 
Attributing \MgII\ emission without accompanying \FeIIs\ to nebular emission, rather than continuum scattering, means that \MgII\ emission alone likely traces \ion{H}{ii} regions within the galaxy and not outflows.
Indeed, galaxies with pure \MgII\ emission profiles have lower SFR surface densities than those with with P-Cygni profiles or \FeIIs\ emission.
The P-Cygni profiles and \FeIIs\ emission signatures likely arise from continuum scattering and fluorescence, since all of these galaxies also have absorption features.
Continuum scattering and fluorescence can produce \FeIIs\ emission either with \MgII\ P-Cygni profiles or with no accompanying \MgII\ emission, in the case of strong resonant trapping. 
Among the emission signatures, \FeIIs\ emission or \MgII\ P-Cygni profiles are therefore the best candidates for tracing outflows.
To confirm that the \FeIIs\ and \MgII\ P-Cygni profiles signatures are associated with galactic outflows, we will need to investigate the kinematics of the absorbing and emitting gas and map the spatial extent, as for the MUSE HDFS galaxy ID\#13 \citep{FinleyH_17a}.

\section{Conclusions}  
\label{sec:conclusions}

Non-resonant \FeIIs\ emission and \MgII\ P-Cygni profiles can potentially trace galactic winds in emission and provide useful constraints on wind models. From the $3.15\arcmin \times 3.15\arcmin$ mosaic of the Hubble Ultra Deep Field (UDF) obtained with the  VLT/MUSE integral field spectrograph, we identify a statistical sample of \NFest\ \FeIIs\ emitters from a sample of 271 \OII\ emitters with reliable redshifts in the range $z=\zmin - \zmax$ down to \fluxlimit. 
From the same parent sample, we identify 50 \MgII\ emitters, with both pure emission and P-Cygni profiles. 
Applying a confidence quality flag (qc~$> 1$), we have 25 \FeIIs\ emitters and 33 \MgII\ emitters, with 9 galaxies that show both emission signatures.

With this sample, we explore the characteristics of galaxies with \FeIIs\ and/or \MgII\ emission. 
Our main results are:

\begin{itemize} 

\item Approximately $10\%$ of galaxies in the redshift range $z=\zmin - \zmax$ have \FeIIs\ or \MgII\ emission with no evidence of an evolution with redshift (Fig.~\ref{fig:zDist}).

\item The \FeIIs\ and \MgII\ emitters follow the galaxy main sequence (Fig.~\ref{fig:Fest:MS}), but show a strong dichotomy. Galaxies below $10^9$ \msun\ (and SFRs of $\lesssim1$ \mpy), have \MgII\ emission without accompanying \FeIIs\ emission, whereas galaxies above $10^{10}$ \msun\ (and SFRs $\gtrsim 10$ \mpy) have \FeIIs\ emission without accompanying \MgII\ emission.  Between these two regimes, galaxies have both \MgII\ and \FeIIs\ emission, typically with \MgII\ P-Cygni profiles.

\item The inclination and size distributions of the \FeIIs\ and \MgII\ emitters are not different from parent samples of \OII\ emitters with similar SFRs, but the size distribution for galaxies with only \MgII\ emission is different from that of galaxies with only \FeIIs\ emission. Consistent with the dichotomy in the SFR-\mstar\ sequence, the galaxies with only \FeIIs\ emission tend to be larger.

\item Splitting the \MgII\ emitter sample by profile type reveals that the galaxies with pure \MgII\ emission profiles have a  star formation rate surface density distribution that is different from galaxies with \MgII\ P-Cygni profiles or \FeIIs\ emission.
The pure \MgII\ emitters have a lower mean value of $-1.1$~\mpy${\rm kpc}^{-2}$, compared to $-0.3$ or $-0.5$~\mpy${\rm kpc}^{-2}$ for \MgII\ P-Cygni profiles or \FeIIs\ emission, and therefore may be less likely to drive outflows. 

\item Representative cases from the UDF-10 field (\ref{fig:udf10-id08}--\ref{fig:udf10-id56}) highlight the progression of \MgII\ spectral signatures from pure emission to P-Cygni profiles to pure absorption, which is likely the result of resonant trapping as the amount of ISM gas and dust increases with stellar mass and SFR.
The representative cases also demonstrate that \FeIIs\ emission consistently occurs with \FeII\ and \MgII\ absorptions, including P-Cygni profiles, whereas pure \MgII\ emission tends to occur without \FeII\ absorption or \FeIIs\ emission.  

\item The UV1 \FeII\,$\lambda2600$ transition and its associated \FeIIs$\lambda2626$ transition are more strongly affected by resonant trapping than the UV2 \FeII\,$\lambda2374$ transition with \FeIIs$\lambda2374$. 
Consequently, the former are more sensitive to dust extinction, which offsets the emission vertically from the photon-conservation line (Fig.~\ref{fig:diagnostic}) and increases as the emission offset increases (Fig.~\ref{fig:offsetAV}).

\end{itemize}


We suggest that different physical mechanisms produce the \FeIIs\ emission and the pure \MgII\ emission. 
Continuum fluorescence, which occurs after absorbing the stellar continuum, gives rise to the \FeIIs\ emission, whereas nebular emission in \ion{H}{ii} regions produces the pure \MgII\ emission. 
In Feltre et al. (in preparation), we will further investigate the physical mechanisms that produce \MgII\ emission with new generation photoionization models to better understand the conditions within the galaxies.

Identifying a statistical sample of individual $z \sim 1$ galaxies with \FeIIs\ emission from MUSE observations creates new opportunities to characterize galactic outflows. 
We will build on the analysis presented in this paper by decomposing the absorption profiles into systemic and blueshifted components to obtain outflow velocities. 
We will also exploit the IFU observations to map the extent of the \FeIIs\ and \MgII\ emission, as in  \citet{FinleyH_17a}.

\begin{acknowledgements}
   We thank for referee for feedback that helped to improve the paper.
   Based on observations collected at the European Organisation for Astronomical Research in the Southern Hemisphere under ESO programs 094.A-0289(B), 095.A-0010(A), 096.A-0045(A) and 096.A-0045(B).
   This work has been carried out thanks to the support of the ANR FOGHAR (ANR-13-BS05-0010-02), the OCEVU Labex (ANR-11-LABX-0060), and the A*MIDEX project (ANR-11-IDEX-0001-02) funded by the ``Investissements d'avenir'' French government program. 
   RB acknowledges support from the ERC advanced grant 339659-MUSICOS.

\end{acknowledgements}

%
%

\bibliographystyle{aa} 
\bibliography{FeII_Biblio} 

\begin{thebibliography}{73}
\expandafter\ifx\csname natexlab\endcsname\relax\def\natexlab#1{#1}\fi

\bibitem[{{Aguirre} {et~al.}(2001){Aguirre}, {Hernquist}, {Schaye}, {Katz},
  {Weinberg}, \& {Gardner}}]{2001ApJ...561..521A}
{Aguirre}, A., {Hernquist}, L., {Schaye}, J., {et~al.} 2001, \apj, 561, 521

\bibitem[{{Arribas} {et~al.}(2014){Arribas}, {Colina}, {Bellocchi}, {Maiolino},
  \& {Villar-Mart{\'{\i}}n}}]{ArribasS_14a}
{Arribas}, S., {Colina}, L., {Bellocchi}, E., {Maiolino}, R., \&
  {Villar-Mart{\'{\i}}n}, M. 2014, \aap, 568, A14

\bibitem[{{Bacon} {et~al.}(2015){Bacon}, {Brinchmann}, {Richard}, {Contini},
  {Drake}, {Franx}, {Tacchella}, {Vernet}, {Wisotzki}, {Blaizot}, {Bouch{\'e}},
  {Bouwens}, {Cantalupo}, {Carollo}, {Carton}, {Caruana}, {Cl{\'e}ment},
  {Dreizler}, {Epinat}, {Guiderdoni}, {Herenz}, {Husser}, {Kamann}, {Kerutt},
  {Kollatschny}, {Krajnovic}, {Lilly}, {Martinsson}, {Michel-Dansac},
  {Patricio}, {Schaye}, {Shirazi}, {Soto}, {Soucail}, {Steinmetz}, {Urrutia},
  {Weilbacher}, \& {de Zeeuw}}]{BaconR_15a}
{Bacon}, R., {Brinchmann}, J., {Richard}, J., {et~al.} 2015, \aap, 575, A75

\bibitem[{{Bacon} {et~al.}(2017){Bacon}, {Conseil}, {Mary},
  {et~al.}}]{Bacon2017}
{Bacon}, R., {Conseil}, D., {Mary}, D., {et~al.} 2017, \aap, in press (MUSE UDF
  SI paper I)

\bibitem[{{Baldry} {et~al.}(2014){Baldry}, {Alpaslan}, \& {Bauer}}]{Baldry_14a}
{Baldry}, I.~K., {Alpaslan}, M., \& {Bauer}, A. 2014, \mnras, 441, 2440

\bibitem[{{Bordoloi} {et~al.}(2017){Bordoloi}, {Fox}, {Lockman}, {Wakker},
  {Jenkins}, {Savage}, {Hernandez}, {Tumlinson}, {Bland-Hawthorn}, \&
  {Kim}}]{BordoloiR_17a}
{Bordoloi}, R., {Fox}, A.~J., {Lockman}, F.~J., {et~al.} 2017, \apj, 834, 191

\bibitem[{{Bordoloi} {et~al.}(2014){Bordoloi}, {Lilly}, {Kacprzak}, \&
  {Churchill}}]{BordoloiR_14a}
{Bordoloi}, R., {Lilly}, S.~J., {Kacprzak}, G.~G., \& {Churchill}, C.~W. 2014,
  \apj, 784, 108

\bibitem[{{Bouch{\'e}} {et~al.}(2010){Bouch{\'e}}, {Dekel}, {Genzel}, {Genel},
  {Cresci}, {F{\"o}rster Schreiber}, {Shapiro}, {Davies}, \&
  {Tacconi}}]{BoucheN_10a}
{Bouch{\'e}}, N., {Dekel}, A., {Genzel}, R., {et~al.} 2010, \apj, 718, 1001

\bibitem[{{Brinchmann} {et~al.}(2004){Brinchmann}, {Charlot}, {White},
  {Tremonti}, {Kauffmann}, {Heckman}, \& {Brinkmann}}]{BrinchmannJ_04a}
{Brinchmann}, J., {Charlot}, S., {White}, S.~D.~M., {et~al.} 2004, \mnras, 351,
  1151

\bibitem[{{Brinchmann} {et~al.}(2017){Brinchmann}, {Inami}, {Bacon},
  {et~al.}}]{Brinchmann2017}
{Brinchmann}, J., {Inami}, H., {Bacon}, R., {et~al.} 2017, \aap, in press (MUSE
  UDF SI paper III)

\bibitem[{{Bruzual} \& {Charlot}(2003)}]{BruzualG_03a}
{Bruzual}, G. \& {Charlot}, S. 2003, \mnras, 344, 1000

\bibitem[{{Calzetti} {et~al.}(2000){Calzetti}, {Armus}, {Bohlin}, {Kinney},
  {Koornneef}, \& {Storchi-Bergmann}}]{CalzettiD_00a}
{Calzetti}, D., {Armus}, L., {Bohlin}, R.~C., {et~al.} 2000, \apj, 533, 682

\bibitem[{{Cameron}(2011)}]{CameronE_11a}
{Cameron}, E. 2011, \pasa, 28, 128

\bibitem[{{Chabrier}(2003)}]{ChabrierG_03a}
{Chabrier}, G. 2003, \apjl, 586, L133

\bibitem[{{Chen} {et~al.}(2010){Chen}, {Tremonti}, {Heckman}, {Kauffmann},
  {Weiner}, {Brinchmann}, \& {Wang}}]{2010AJ....140..445C}
{Chen}, Y.-M., {Tremonti}, C.~A., {Heckman}, T.~M., {et~al.} 2010, \aj, 140,
  445

\bibitem[{{Chisholm} {et~al.}(2015){Chisholm}, {Tremonti}, {Leitherer}, {Chen},
  {Wofford}, \& {Lundgren}}]{ChisholmJ_15a}
{Chisholm}, J., {Tremonti}, C.~A., {Leitherer}, C., {et~al.} 2015, \apj, 811,
  149

\bibitem[{{Coil} {et~al.}(2011){Coil}, {Weiner}, {Holz}, {Cooper}, {Yan}, \&
  {Aird}}]{CoilA_11a}
{Coil}, A.~L., {Weiner}, B.~J., {Holz}, D.~E., {et~al.} 2011, \apj, 743, 46

\bibitem[{{Dutta} {et~al.}(2017){Dutta}, {Srianand}, {Gupta}, {Joshi},
  {Petitjean}, {Noterdaeme}, {Ge}, \& {Krogager}}]{DuttaR_17a}
{Dutta}, R., {Srianand}, R., {Gupta}, N., {et~al.} 2017, \mnras, 465, 4249

\bibitem[{{Erb} {et~al.}(2012){Erb}, {Quider}, {Henry}, \& {Martin}}]{ErbD_12a}
{Erb}, D.~K., {Quider}, A.~M., {Henry}, A.~L., \& {Martin}, C.~L. 2012, \apj,
  759, 26

\bibitem[{{Finlator} \& {Dav{\'e}}(2008)}]{2008MNRAS.385.2181F}
{Finlator}, K. \& {Dav{\'e}}, R. 2008, \mnras, 385, 2181

\bibitem[{{Finley} {et~al.}(2017){Finley}, {Bouch{\'e}}, {Contini}, {Epinat},
  {Bacon}, {Brinchmann}, {Cantalupo}, {Erroz-Ferrer}, {Marino}, {Maseda},
  {Richard}, {Schroetter}, {Verhamme}, {Weilbacher}, {Wendt}, \&
  {Wisotzki}}]{FinleyH_17a}
{Finley}, H., {Bouch{\'e}}, N., {Contini}, T., {et~al.} 2017, \aap, 605, A118

\bibitem[{{Ford} {et~al.}(2016){Ford}, {Werk}, {Dav{\'e}}, {Tumlinson},
  {Bordoloi}, {Katz}, {Kollmeier}, {Oppenheimer}, {Peeples}, {Prochaska}, \&
  {Weinberg}}]{2016MNRAS.459.1745F}
{Ford}, A.~B., {Werk}, J.~K., {Dav{\'e}}, R., {et~al.} 2016, \mnras, 459, 1745

\bibitem[{{Fox} {et~al.}(2015){Fox}, {Bordoloi}, {Savage}, {Lockman},
  {Jenkins}, {Wakker}, {Bland-Hawthorn}, {Hernandez}, {Kim}, {Benjamin},
  {Bowen}, \& {Tumlinson}}]{FoxA_15a}
{Fox}, A.~J., {Bordoloi}, R., {Savage}, B.~D., {et~al.} 2015, \apjl, 799, L7

\bibitem[{{Genzel} {et~al.}(2011){Genzel}, {Newman}, {Jones}, {F{\"o}rster
  Schreiber}, {Shapiro}, {Genel}, {Lilly}, {Renzini}, {Tacconi}, {Bouch{\'e}},
  {Burkert}, {Cresci}, {Buschkamp}, {Carollo}, {Ceverino}, {Davies}, {Dekel},
  {Eisenhauer}, {Hicks}, {Kurk}, {Lutz}, {Mancini}, {Naab}, {Peng},
  {Sternberg}, {Vergani}, \& {Zamorani}}]{GenzelR_11a}
{Genzel}, R., {Newman}, S., {Jones}, T., {et~al.} 2011, \apj, 733, 101

\bibitem[{{Grimes} {et~al.}(2005){Grimes}, {Heckman}, {Strickland}, \&
  {Ptak}}]{GrimesJ_2005a}
{Grimes}, J.~P., {Heckman}, T., {Strickland}, D., \& {Ptak}, A. 2005, \apj,
  628, 187

\bibitem[{{Harikane} {et~al.}(2014){Harikane}, {Ouchi}, {Yuma}, {Rauch},
  {Nakajima}, \& {Ono}}]{HarikaneY_14a}
{Harikane}, Y., {Ouchi}, M., {Yuma}, S., {et~al.} 2014, \apj, 794, 129

\bibitem[{{Heckman}(2002)}]{HeckmanT_02a}
{Heckman}, T.~M. 2002, in Astronomical Society of the Pacific Conference
  Series, Vol. 254, Extragalactic Gas at Low Redshift, ed. J.~S. {Mulchaey} \&
  J.~T. {Stocke}, 292

\bibitem[{{Heckman} {et~al.}(2015){Heckman}, {Alexandroff}, {Borthakur},
  {Overzier}, \& {Leitherer}}]{HeckmanT_15a}
{Heckman}, T.~M., {Alexandroff}, R.~M., {Borthakur}, S., {Overzier}, R., \&
  {Leitherer}, C. 2015, \apj, 809, 147

\bibitem[{{Hinton} {et~al.}(2016){Hinton}, {Davis}, {Lidman}, {Glazebrook}, \&
  {Lewis}}]{Hinton_16a}
{Hinton}, S.~R., {Davis}, T.~M., {Lidman}, C., {Glazebrook}, K., \& {Lewis},
  G.~F. 2016, Astronomy and Computing, 15

\bibitem[{{Ho} {et~al.}(2016){Ho}, {Medling}, {Bland-Hawthorn}, {Groves},
  {Kewley}, {Kobayashi}, {Dopita}, {Leslie}, {Sharp}, {Allen}, {Bourne},
  {Bryant}, {Cortese}, {Croom}, {Dunne}, {Fogarty}, {Goodwin}, {Green},
  {Konstantopoulos}, {Lawrence}, {Lorente}, {Owers}, {Richards}, {Sweet},
  {Tescari}, \& {Valiante}}]{HoI_2016a}
{Ho}, I.-T., {Medling}, A.~M., {Bland-Hawthorn}, J., {et~al.} 2016, \mnras,
  457, 1257

\bibitem[{{Inami} {et~al.}(2017){Inami}, {Bacon}, {Brinchmann},
  {et~al.}}]{Inami2017}
{Inami}, H., {Bacon}, R., {Brinchmann}, J., {et~al.} 2017, \aap, in press (MUSE
  UDF SI paper II)

\bibitem[{{Karim} {et~al.}(2011){Karim}, {Schinnerer},
  {Mart{\'{\i}}nez-Sansigre}, {Sargent}, {van der Wel}, {Rix}, {Ilbert},
  {Smol{\v c}i{\'c}}, {Carilli}, {Pannella}, {Koekemoer}, {Bell}, \&
  {Salvato}}]{KarimA_11a}
{Karim}, A., {Schinnerer}, E., {Mart{\'{\i}}nez-Sansigre}, A., {et~al.} 2011,
  \apj, 730, 61

\bibitem[{{Kewley} {et~al.}(2004){Kewley}, {Geller}, \& {Jansen}}]{KewleyL_04a}
{Kewley}, L.~J., {Geller}, M.~J., \& {Jansen}, R.~A. 2004, \aj, 127, 2002

\bibitem[{{Kornei} {et~al.}(2013){Kornei}, {Shapley}, {Martin}, {Coil}, {Lotz},
  \& {Weiner}}]{KorneiK_13a}
{Kornei}, K.~A., {Shapley}, A.~E., {Martin}, C.~L., {et~al.} 2013, \apj, 774,
  50

\bibitem[{{Kriek} {et~al.}(2009){Kriek}, {van Dokkum}, {Labb{\'e}}, {Franx},
  {Illingworth}, {Marchesini}, \& {Quadri}}]{KriekM_09a}
{Kriek}, M., {van Dokkum}, P.~G., {Labb{\'e}}, I., {et~al.} 2009, \apj, 700,
  221

\bibitem[{{Lehnert} \& {Heckman}(1995)}]{LehnertM_1995a}
{Lehnert}, M.~D. \& {Heckman}, T.~M. 1995, \apjs, 97, 89

\bibitem[{{Lehnert} \& {Heckman}(1996)}]{LehnertM_1996a}
{Lehnert}, M.~D. \& {Heckman}, T.~M. 1996, \apj, 462, 651

\bibitem[{{Lehnert} {et~al.}(1999){Lehnert}, {Heckman}, \&
  {Weaver}}]{LehnertM_99a}
{Lehnert}, M.~D., {Heckman}, T.~M., \& {Weaver}, K.~A. 1999, \apj, 523, 575

\bibitem[{{Lilly} {et~al.}(2013){Lilly}, {Carollo}, {Pipino}, {Renzini}, \&
  {Peng}}]{2013ApJ...772..119L}
{Lilly}, S.~J., {Carollo}, C.~M., {Pipino}, A., {Renzini}, A., \& {Peng}, Y.
  2013, \apj, 772, 119

\bibitem[{{Martin}(1999)}]{MartinC_99a}
{Martin}, C.~L. 1999, \apj, 513, 156

\bibitem[{{Martin} {et~al.}(2012){Martin}, {Shapley}, {Coil}, {Kornei},
  {Bundy}, {Weiner}, {Noeske}, \& {Schiminovich}}]{MartinC_12a}
{Martin}, C.~L., {Shapley}, A.~E., {Coil}, A.~L., {et~al.} 2012, \apj, 760, 127

\bibitem[{{Martin} {et~al.}(2013){Martin}, {Shapley}, {Coil}, {Kornei},
  {Murray}, \& {Pancoast}}]{MartinC_13a}
{Martin}, C.~L., {Shapley}, A.~E., {Coil}, A.~L., {et~al.} 2013, \apj, 770, 41

\bibitem[{{Mitra} {et~al.}(2017){Mitra}, {Dav{\'e}}, {Simha}, \&
  {Finlator}}]{MitraS_17a}
{Mitra}, S., {Dav{\'e}}, R., {Simha}, V., \& {Finlator}, K. 2017, \mnras, 464,
  2766

\bibitem[{{Newman} {et~al.}(2012){Newman}, {Genzel}, {F{\"o}rster-Schreiber},
  {Shapiro Griffin}, {Mancini}, {Lilly}, {Renzini}, {Bouch{\'e}}, {Burkert},
  {Buschkamp}, {Carollo}, {Cresci}, {Davies}, {Eisenhauer}, {Genel}, {Hicks},
  {Kurk}, {Lutz}, {Naab}, {Peng}, {Sternberg}, {Tacconi}, {Vergani}, {Wuyts},
  \& {Zamorani}}]{NewmanS_12a}
{Newman}, S.~F., {Genzel}, R., {F{\"o}rster-Schreiber}, N.~M., {et~al.} 2012,
  \apj, 761, 43

\bibitem[{{Oppenheimer} \& {Dav{\'e}}(2008)}]{2008MNRAS.387..577O}
{Oppenheimer}, B.~D. \& {Dav{\'e}}, R. 2008, \mnras, 387, 577

\bibitem[{{Peng} {et~al.}(2010){Peng}, {Ho}, {Impey}, \& {Rix}}]{PengC_10a}
{Peng}, C.~Y., {Ho}, L.~C., {Impey}, C.~D., \& {Rix}, H.-W. 2010, \aj, 139,
  2097

\bibitem[{{Prochaska} {et~al.}(2011){Prochaska}, {Kasen}, \&
  {Rubin}}]{ProchaskaJ_11a}
{Prochaska}, J.~X., {Kasen}, D., \& {Rubin}, K. 2011, \apj, 734, 24

\bibitem[{{Rafelski} {et~al.}(2015){Rafelski}, {Teplitz}, {Gardner}, {Coe},
  {Bond}, {Koekemoer}, {Grogin}, {Kurczynski}, {McGrath}, {Bourque}, {Atek},
  {Brown}, {Colbert}, {Codoreanu}, {Ferguson}, {Finkelstein}, {Gawiser},
  {Giavalisco}, {Gronwall}, {Hanish}, {Lee}, {Mehta}, {de Mello},
  {Ravindranath}, {Ryan}, {Scarlata}, {Siana}, {Soto}, \&
  {Voyer}}]{RafelskiM_15a}
{Rafelski}, M., {Teplitz}, H.~I., {Gardner}, J.~P., {et~al.} 2015, \aj, 150, 31

\bibitem[{{Rubin} {et~al.}(2014){Rubin}, {Prochaska}, {Koo}, {Phillips},
  {Martin}, \& {Winstrom}}]{RubinK_14a}
{Rubin}, K.~H.~R., {Prochaska}, J.~X., {Koo}, D.~C., {et~al.} 2014, \apj, 794,
  156

\bibitem[{{Rubin} {et~al.}(2011){Rubin}, {Prochaska}, {M{\'e}nard}, {Murray},
  {Kasen}, {Koo}, \& {Phillips}}]{RubinK_11a}
{Rubin}, K.~H.~R., {Prochaska}, J.~X., {M{\'e}nard}, B., {et~al.} 2011, \apj,
  728, 55

\bibitem[{{Rubin} {et~al.}(2010){Rubin}, {Weiner}, {Koo}, {Martin},
  {Prochaska}, {Coil}, \& {Newman}}]{RubinK_10a}
{Rubin}, K.~H.~R., {Weiner}, B.~J., {Koo}, D.~C., {et~al.} 2010, \apj, 719,
  1503

\bibitem[{{Rupke} \& {Veilleux}(2013)}]{RupkeD_13a}
{Rupke}, D.~S.~N. \& {Veilleux}, S. 2013, \apj, 768, 75

\bibitem[{{Rupke} \& {Veilleux}(2015)}]{2015ApJ...801..126R}
{Rupke}, D.~S.~N. \& {Veilleux}, S. 2015, \apj, 801, 126

\bibitem[{{Salpeter}(1955)}]{SalpeterE_55a}
{Salpeter}, E.~E. 1955, \apj, 121, 161

\bibitem[{{Scarlata} \& {Panagia}(2015)}]{ScarlataC_15a}
{Scarlata}, C. \& {Panagia}, N. 2015, \apj, 801, 43

\bibitem[{{Schreiber} {et~al.}(2015){Schreiber}, {Pannella}, {Elbaz},
  {B{\'e}thermin}, {Inami}, {Dickinson}, {Magnelli}, {Wang}, {Aussel}, {Daddi},
  {Juneau}, {Shu}, {Sargent}, {Buat}, {Faber}, {Ferguson}, {Giavalisco},
  {Koekemoer}, {Magdis}, {Morrison}, {Papovich}, {Santini}, \&
  {Scott}}]{SchreiberC_15a}
{Schreiber}, C., {Pannella}, M., {Elbaz}, D., {et~al.} 2015, \aap, 575, A74

\bibitem[{{Sharma} {et~al.}(2016){Sharma}, {Theuns}, {Frenk}, {Bower}, {Crain},
  {Schaller}, \& {Schaye}}]{SharmaM_16a}
{Sharma}, M., {Theuns}, T., {Frenk}, C., {et~al.} 2016, \mnras, 458, L94

\bibitem[{{Silk} \& {Mamon}(2012)}]{2012RAA....12..917S}
{Silk}, J. \& {Mamon}, G.~A. 2012, Research in Astronomy and Astrophysics, 12,
  917

\bibitem[{{Soto} \& {Martin}(2012)}]{SotoK_12a}
{Soto}, K.~T. \& {Martin}, C.~L. 2012, \apjs, 203, 3

\bibitem[{{Steidel} {et~al.}(2010){Steidel}, {Erb}, {Shapley}, {Pettini},
  {Reddy}, {Bogosavljevi{\'c}}, {Rudie}, \& {Rakic}}]{SteidelC_10a}
{Steidel}, C.~C., {Erb}, D.~K., {Shapley}, A.~E., {et~al.} 2010, \apj, 717, 289

\bibitem[{Strickland \& Heckman(2009)}]{StricklandD_09a}
Strickland, D.~K. \& Heckman, T.~M. 2009, 29

\bibitem[{{Strickland} {et~al.}(2004){Strickland}, {Heckman}, {Colbert},
  {Hoopes}, \& {Weaver}}]{StricklandD_04a}
{Strickland}, D.~K., {Heckman}, T.~M., {Colbert}, E.~J.~M., {Hoopes}, C.~G., \&
  {Weaver}, K.~A. 2004, \apjs, 151, 193

\bibitem[{{Strickland} \& {Stevens}(1999)}]{StricklandD_99a}
{Strickland}, D.~K. \& {Stevens}, I.~R. 1999, \mnras, 306, 43

\bibitem[{{Tang} {et~al.}(2014){Tang}, {Giavalisco}, {Guo}, \&
  {Kurk}}]{TangY_14a}
{Tang}, Y., {Giavalisco}, M., {Guo}, Y., \& {Kurk}, J. 2014, \apj, 793, 92

\bibitem[{{Tremonti} {et~al.}(2004){Tremonti}, {Heckman}, {Kauffmann},
  {Brinchmann}, {Charlot}, {White}, {Seibert}, {Peng}, {Schlegel}, {Uomoto},
  {Fukugita}, \& {Brinkmann}}]{TremontiC_04a}
{Tremonti}, C.~A., {Heckman}, T.~M., {Kauffmann}, G., {et~al.} 2004, \apj, 613,
  898

\bibitem[{{van der Wel} {et~al.}(2012){van der Wel}, {Bell}, {H{\"a}ussler},
  {McGrath}, {Chang}, {Guo}, {McIntosh}, {Rix}, {Barden}, {Cheung}, {Faber},
  {Ferguson}, {Galametz}, {Grogin}, {Hartley}, {Kartaltepe}, {Kocevski},
  {Koekemoer}, {Lotz}, {Mozena}, {Peth}, \& {Peng}}]{VanderWelA_12a}
{van der Wel}, A., {Bell}, E.~F., {H{\"a}ussler}, B., {et~al.} 2012, \apjs,
  203, 24

\bibitem[{{Veilleux} {et~al.}(2005){Veilleux}, {Cecil}, \&
  {Bland-Hawthorn}}]{VeilleuxS_05a}
{Veilleux}, S., {Cecil}, G., \& {Bland-Hawthorn}, J. 2005, \araa, 43, 769

\bibitem[{{Veilleux} {et~al.}(2003){Veilleux}, {Shopbell}, {Rupke},
  {Bland-Hawthorn}, \& {Cecil}}]{VeilleuxS_03a}
{Veilleux}, S., {Shopbell}, P.~L., {Rupke}, D.~S., {Bland-Hawthorn}, J., \&
  {Cecil}, G. 2003, \aj, 126, 2185

\bibitem[{{Weiner} {et~al.}(2009){Weiner}, {Coil}, {Prochaska}, {Newman},
  {Cooper}, {Bundy}, {Conselice}, {Dutton}, {Faber}, {Koo}, {Lotz}, {Rieke}, \&
  {Rubin}}]{WeinerB_09a}
{Weiner}, B.~J., {Coil}, A.~L., {Prochaska}, J.~X., {et~al.} 2009, \apj, 692,
  187

\bibitem[{{Westmoquette} {et~al.}(2012){Westmoquette}, {Clements}, {Bendo}, \&
  {Khan}}]{WestmoquetteM_12a}
{Westmoquette}, M.~S., {Clements}, D.~L., {Bendo}, G.~J., \& {Khan}, S.~A.
  2012, \mnras, 424, 416

\bibitem[{Whitaker {et~al.}(2014)Whitaker, Franx, Leja, van Dokkum, Henry,
  Skelton, Fumagalli, Momcheva, Brammer, Labb{\'{e}}, Nelson, \&
  Rigby}]{WhitakerK_14a}
Whitaker, K.~E., Franx, M., Leja, J., {et~al.} 2014, The Astrophysical Journal,
  795, 104

\bibitem[{{Zahid} {et~al.}(2014){Zahid}, {Dima}, {Kudritzki}, {Kewley},
  {Geller}, {Hwang}, {Silverman}, \& {Kashino}}]{ZahidJ_14b}
{Zahid}, H.~J., {Dima}, G.~I., {Kudritzki}, R.-P., {et~al.} 2014, \apj, 791,
  130

\bibitem[{{Zhu} {et~al.}(2015){Zhu}, {Comparat}, {Kneib}, {Delubac},
  {Raichoor}, {Dawson}, {Newman}, {Y{\`e}che}, {Zhou}, \&
  {Schneider}}]{ZhuG_15a}
{Zhu}, G.~B., {Comparat}, J., {Kneib}, J.-P., {et~al.} 2015, \apj, 815, 48

\end{thebibliography}

\end{document}